\documentclass[useAMS,usenatbib,usegraphicx]{mn2e}

\usepackage{times}
\usepackage{epsf}
\begin{document}
\title[Large-Scale Structure Around Distant MACS Clusters]{Probing the Large-Scale Structure Around the Most Distant Galaxy Clusters from the Massive Cluster Survey}

\author[J. S.\ Kartaltepe et al.]{Jeyhan S. Kartaltepe\thanks{E-mail: jeyhan@ifa.hawaii.edu}, Harald Ebeling, C.J. Ma, \& David Donovan \\ Institute for Astronomy, University of Hawaii, 2680 Woodlawn  
Drive, Honolulu, HI 96822, USA}

\maketitle

\begin{abstract}
We present maps of the cosmic large-scale structure around the twelve most distant galaxy clusters from the Massive Cluster Survey (MACS) as traced by the projected surface density of galaxies on the cluster red sequence. Taken with the Suprime-Cam wide-field camera on the Subaru telescope, the images used in this study cover a $27\times 27$ arcmin$^2$ area around each cluster, corresponding to $10 \times 10$ Mpc$^2$ at the median redshift of $z=0.55$ of our sample. We directly detect satellite clusters and filaments extending over the full size of our imaging data in the majority of the clusters studied, supporting the picture of mass accretion via infall along filaments suggested by numerical simulations of the growth of clusters and the evolution of large-scale structure. A comparison of the galaxy distribution near the cluster cores with the X-ray surface brightness as observed with Chandra reveals, in several cases, significant offsets between the gas and galaxy distribution, indicative of ongoing merger events. The respective systems are ideally suited for studies of the dynamical properties of gas, galaxies, and dark matter. In addition, the large-scale filaments viewed at high contrast in these MACS clusters are prime targets for the direct detection and study of the warm-hot intergalactic medium (WHIM).
\end{abstract}


\section{INTRODUCTION}

As the largest gravitationally bound constituents of matter, galaxy clusters provide important clues about the distribution and fundamental properties of galaxies, gas, and dark matter. Studying their characteristics over a range of redshifts can yield crucial insights into the overall structure and evolution of the cosmos. Theoretical considerations and the results of numerical simulations (e.g.\ Colberg et al.\ 2000; Evrard et al.\ 2002) predict that clusters grow through accretion and merger events and reside within filamentary structures extending up to $10/h$ Mpc in size. While growth through cluster mergers is a relatively well studied phenomenon, the existence and relevance of filaments for cluster and galaxy evolution remains much less well understood, primarily because of observational challenges created by the very large scales and low densities involved.

Numerical simulations (e.g.\ Cen \& Ostriker 1999, Dav\'{e} et al.\ 1999) suggest that a large fraction of the baryons in the local universe takes the form of highly diffuse gas at a temperature of $10^{5} - 10^{7} K$ -- the so-called ``Warm-Hot Intergalactic Medium" (WHIM). The temperature and diffuse nature of the WHIM makes it very difficult to detect directly, the currently most promising techniques being very deep observations at extreme ultraviolet (EUV) and soft X-ray wavelengths, to detect high-ionization metal absorption lines and diffuse bremsstrahlung emission, respectively (e.g. Soltan, Freyberg \& Hasinger 2005, Nicastro et al.\ 2005). Since simulations also predict that the WHIM is distributed along large-scale filaments around galaxy clusters (Dav\'{e} et al.\ 2001) and may be detectable in observations of the Sunyaev-Zel'dovich Effect (Hallman et al.\ 2007), promising targets for the direct detection and subsequent study of the WHIM could be unveiled by a systematic survey of the large-scale structure surrounding massive galaxy clusters.

Large optical surveys of the nearby universe such as 2dF and SDSS have allowed detailed investigations of the galaxies within the vast cosmic web consisting of clusters, superclusters, voids, walls, and filamentary structures (e.g.\ Peacock et al.\ 2001; Zandivarez, Merch\'{a}n \& Padilla 2003; Erdo\^{g}du et al.\ 2004; Einasto et al.\ 2007). Because of the limited depth of the underlying imaging and spectroscopic data obtained by these surveys, studies of low-density environments have focused on structures at relatively low redshift ($z \sim 0.1-0.2$). A search for filamentary structure around the vertices of the ``cosmic web'' would, however, be particularly rewarding at higher redshifts where the contrast is greatly enhanced by the smaller angular size of such features. In addition, observations at higher redshift would establish an important baseline for evolutionary studies of this environment. Galaxy clusters at higher redshifts are expected to be less evolved and reside within the densest regions of the precursors of the structure seen locally. A large fraction of the matter, dark and luminous, in these structures will eventually become part of virialized cluster cores through merging and infall along the filaments. Hence a multitude of astrophysical topics can be addressed by studying the large-scale surroundings of distant clusters, including the impact of environment on star formation in galaxies, the relative importance of gradual infall along filaments vs. merging for both galaxy and cluster evolution, and the existence and properties of the WHIM.

The study described in the following looks for evidence of filamentary structure in a complete sample of twelve moderately distant, very massive galaxy clusters in order to establish observationally the frequency and prevalance of such structure around the densest nodes of the cosmic web. By targeting galaxy clusters at intermediate redshift ($z \sim 0.5$), we observe filaments on angular scales of typically 10 arcmin, which makes them accessible to efficient observational sampling in wavebands from the optical to X-rays. The high mass, intermediate redshift, and compact angular scale of these systems make them particularly promising targets for the detection of WHIM emission. Previous studies of the large-scale environment of clusters at similar redshift (e.g. Kodama et al.\ 2001) have either focused on a single object or on systems of relatively low mass. Since more massive clusters are connected to more, and denser, filaments than less massive systems (Colberg, Krughoff, \& Connolly
2005, Colberg 2007), a systematic survey of the environment of a statistically complete sample of the most massive clusters at these redshifts is advantageous in the context of the science goals outlined above.

The structure of this paper is as follows. In \S2 we describe the selection of our cluster sample. \S3 describes the observations and data reduction as well as the methods used in our analysis of the data. We present and discuss our results in \S4. \S5 contains our conclusions and identifies areas for future work. Throughout the paper we assume a cosmology with $ \Omega_M = 0.3, \Omega_\Lambda = 0.7$, and $h_0 = 0.7$.

\section{Cluster Sample Selection}

The galaxy clusters used in this study are a subset of the systems detected in the Massive Cluster Survey (MACS), a survey designed with the aim of finding the distant counterparts to the most X-ray luminous nearby clusters (Ebeling, Edge, \& Henry 2001). We here focus on the complete sample of the twelve most distant MACS clusters at $z>0.5$ as published by Ebeling et al.\ (2007). As the most extreme representatives of a sample of already extraordinarily massive clusters (typically $10^{15} M_{\odot}$), these most distant MACS clusters have been high-priority targets of an ongoing comprehensive multiwavelength follow-up campaign, ensuring that wide-field optical and X-ray data are available for all 12 clusters. At the sample's median redshift of $z=0.55$ the expected scale of the large-scale structure we aim to study is ideally matched to the field of view of modern wide-field cameras, eliminating the need for mosaicking. A pilot study of one of these clusters (MACS\,J0717.5+3745) unveiled a $\sim 4$ Mpc long filament (Ebeling, Barrett, \& Donovan 2004), a discovery that provided the impetus for this comprehensive study of the environment surrounding similarly massive clusters in order to determine whether such filaments are rare for these systems or whether they are the norm.

\section{Data Acquisition and Analysis}

\subsection{Optical}

As part of our comprehensive observational follow-up campaign of MACS clusters we imaged all 12 clusters of this sample with the SuprimeCam wide-field imager on the Subaru 8.2m telescope on Mauna Kea (Miyazaki et al.\ 2002). Covering a field of view of approximately $34\times 27$ arcmin$^2$ (corresponding to $13\times 10 $ Mpc$^2$ at the median redshift of our cluster targets), SuprimeCam is ideally suited to study the large-scale environment of massive clusters. We here use data obtained in the $V$ and $R$ broadband filters, taking advantage of the fact that these passbands bracket the D4000 \AA\ break at the redshifts of our clusters and thus allow an efficient selection of early-type cluster members using the cluster red sequence.

The imaging data used here were obtained in several observing runs between December 2000 and November 
2005 in variable, but usually near-photometric, conditions. The seeing ranged from 0.7 to 1.2 arcsec as measured from the final stacked images. The integration times were, typically, 48 minutes in $R$ and 36 min in $V$.  All data were reduced employing standard techniques which have, however, been adapted and refined specifically for the analysis of SuprimeCam data; details of the procedure are given by Donovan (2007).

In order to allow the measurement of robust $V-R$ colors for all objects within the field of view we ensured that the $V$ and $R$ band images for each cluster had the same effective spatial resolution by blurring the image obtained in better conditions until its seeing, as measured for several hundred stars in the field, matched the one obtained in poorer seeing. The seeing in these seeing-matched images ranged from 0.9 to 1.2 arcsec for the 12 clusters of our sample. For the clusters falling within the footprint of the Sloan Digital Sky Survey (SDSS), the photometric zeropoints were derived using the stars in the SDSS catalogue\footnote{The SDSS photometry has been transformed into the Johnson-Cousin system using the equation of Lupton (2005) on the SDSS homepage: http://www.sdss.org/dr5/algorithms/sdssUBVRITransform.html}. For those cluster fields not covered by the SDSS we obtained cross-calibration in the form of 3-sec. exposures on nearby SDSS fields.

Object catalogues for each cluster were then created with SExtractor Version 2.4.3 (Bertin \& Arnouts 1996) in ``dual image'' mode with the R-band image as the reference detection image. We separated stars from galaxies using fits to the stellar sequence in both the magnitude vs. peak surface-brightness distribution and the magnitude vs. half-light radius distribution. We fit a straight line to the stellar locus and consider objects farther than 3$\sigma$ from the best fit to be galaxies. For the galaxies thus selected we created $V-R$ vs $R$ color-magnitude diagrams using fixed apertures of 1 arcsec radius for the color measurement, combined with total isophotal magnitudes in the $R$ band as returned by SExtractor's {\sc mag\_auto} parameter.

We then co-added the color-magnitude diagrams for all of the clusters after shifting the various data sets along the $V-R$ axis such that, for each cluster, the median color of the fifty brightest cluster galaxies matched the respective value for a reference cluster near the median redshift of our sample (MACS\,J0717.5+3745, $z=0.545$). The $R$ magnitudes were corrected to the same median cluster redshift of $z=0.55$ using the distance modulus and a K-correction obtained with a template appropriate for a red sequence galaxy and generated using a suitable model from Bruzual \& Charlot (2003) with a Salpeter initial mass function, solar metallicity, and assuming a formation redshift of $z=5.0$. The resulting stacked color-magnitude diagram is shown in Fig. 1. The color-selected galaxy sample used in our analysis was obtained by performing a linear fit to the red sequence in this stacked color-magnitude diagram down to an $R$ magnitude of 22.5. We then selected all galaxies within 2$\sigma$ (as defined as the gaussian width of the galaxies around the best-fit line) of this best-fit red sequence down to an $R$ magnitude of 23.5 as shown by the orange polygon in Fig. 1. This magnitude limit was chosen by measuring the global $90\%$ completeness limit for all galaxies in the entire sample\footnote{For the most distant cluster of our sample, MACS\,J0744.8+3927, $90\%$ completeness is already reached at $R\sim 22$ (after correcting to $z=0.55$); we hence applied this more stringent cut for this one cluster.} and corresponds to $0.03 L^\star$ at $z=0.55$. For the purposes of this work we consider all of these red-sequence selected galaxies to be cluster members and tracers of the densest and most highly evolved part of the galaxy population of our cluster targets and their environments. Further details for each cluster are given in Table 1.

\begin{figure}
\includegraphics[width=3.5in]{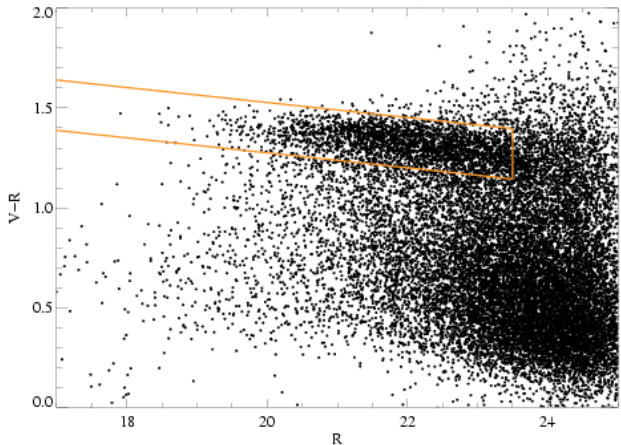}
\caption{Stacked color-magnitude diagram for our sample, after correcting  magnitudes to a nominal cluster redshift of $z=0.55$ and shifting $V-R$ colors such that the median color of the first to fortyninth-ranked galaxies in all cluster red sequences match. The orange polygon marks the region within which
  we consider galaxies to be cluster members.}
\end{figure}

\begin{table*}
\begin{minipage}{5.5in}
  \caption{MACS Clusters at $z>0.5$} 
  \label{Clusters}
   \begin{tabular}{lccccc}
  \hline
  Cluster Name$^a$ & Redshift & \# of Galaxies$^b$ & 
  $t_{\rm exp}$ $(V, R)$ (s)$^c$ & Date Observed $(V, R)$ & $t_{\rm exp}$ (X-ray) (ks)$^c$ \\
  \hline
    MACS\,J0018.5+1626 & 0.5476 & 1515 & 2160, 2880 & Sep. 03, Dec. 02 & 70.1 \\
    MACS\,J0025.4-1222 & 0.5840 & 1170 & 2160, 2880 & Sep. 03, Dec. 02 & 27.4 \\
    MACS\,J0257.1-2335 & 0.5058 &  883 & 2160, 5280 & Dec. 02, Dec. 00 & 20.5 \\
    MACS\,J0454.1-0300 & 0.5377 &  997 & 2160, 3240 & Nov. 05, Mar. 05 & 16.9 \\
    MACS\,J0647.7+7015 & 0.5912 & 1478 & 2160, 2880 & Sep. 03, Feb. 04 & 22.1 \\
    MACS\,J0717.5+3745 & 0.5457 & 1872 & 2160, 2880 & Dec. 02, Dec. 00 & 61.7 \\
    MACS\,J0744.8+3927 & 0.6984 & 1231 & 1440, 2880 & Apr. 03, Dec. 02 & 53.2 \\
    MACS\,J0911.2+1746 & 0.5041 &  849 & 2160, 2880 & Apr. 03, Apr. 03 & 26.7 \\
    MACS\,J1149.5+2223 & 0.5445 & 1589 & 2160, 2880 & Apr. 03, Apr. 03 & 22.0 \\
    MACS\,J1423.8+2404 & 0.5432 & 1093 & 2160, 2400 & Jun. 02, Jun. 02 & 20.2 \\
    MACS\,J2129.4-0741 & 0.5347 & 1108 & 1440, 2880 & Jun. 02, Jun. 01 & 21.7 \\
    MACS\,J2214.9-1359 & 0.5030 & 1462 & 2160, 2880 & Jul. 07, Jul. 03 & 20.1 \\
	\hline
    \end{tabular}
    \footnotetext{$^a$ For coordinates and other cluster properties, see Ebeling et al.\ 2007 } 
    \footnotetext{$^b$ Number of galaxies selected along the red sequence using the method described in section 3.1}
    \footnotetext{$^c$ Exposure times of the optical and X-ray data in seconds and kiloseconds, respectively} 
    \end{minipage}
\end{table*}

\subsection{X-ray Observations}

The X-ray data were obtained with the ACIS-I instrument (the imager of the Advanced CCD Imaging Spectrometer) aboard the Chandra X-Ray Observatory between January 2001 and December 2005. The exposure time ranged from 16 to 70 ks. All observations were conducted in VFaint mode and used chips 1, 2, 3, 4, and 6 of the ACIS CCD array. The data were reduced using standard procedures of the CIAO 3.0.2 software suite. The X-ray images in the 0.5--7 keV band were smoothed using the adaptive-kernel smoothing algorithm, {\sc Asmooth} (Ebeling, White, \& Rangarajan 2006) to produce the contour maps discussed in the next section.

\section{RESULTS AND DISCUSSION}

Figure 2 shows, for each of the twelve clusters in our sample, contours of the adaptively smoothed, projected surface density of the red-sequence selected galaxies as obtained with {\sc Asmooth}. By design, all features within the bold outline are significant at the $3\sigma$ confidence level. Figure 3 shows the same data in a scaled representation, but only for the cluster cores and with X-ray surface brightness countours overlaid.

\subsection{Optical Morphology}

All but three of the twelve clusters show evidence for at least one satellite cluster, the exceptions being MACS\,J0911.2$+$1746 (which, however, consists of two close subclusters), MACS\,J1423.8+2404, and MACS\,J2129.4$-$0741. The prominent satellite cluster almost 10 arcmin southwest of the core of MACS\,J0018+1626 (A.K.A. Cl0016+16, Spinrad 1980) has been previously identified (Connolly et al.\ 1996) but the filament connecting the two systems is reported here for the first time. More than half of all systems feature more than one apparent satellite (MACS\,J0018+1626, MACS\,J0025.4--1222, MACS\,J0257.1--2335, MACS\,J0454.1$-$0300, MACS\,J0647.7+7015, MACS\,J0717.5+3745, MACS\,J2214.9$-$1359) and an equal number show well-defined filamentary structure (MACS\,J0018+1626, MACS\,J0257.1--2335, MACS\,J0454.1$-$0300, MACS\,J0647.7+7015, MACS\,J0717.5+3745, MACS\,J2129.4$-$0741, MACS\,J2214.9$-$1359). The filament extending to the southeast of the main cluster in MACS\,J0717.5+4735 was previously identified by Ebeling et al.\ (2004) who point out that its orientation matches that of the merger axis of the cluster proper, suggesting that the system has grown for many Gyr via infall of matter along this filament. We find similar alignments between filaments and the galaxy-density contours of the cluster core for almost all other systems exhibiting filamentary structure (seven out of twelve as listed above), providing strong observational evidence that the growth of most clusters occurs along a single preferred direction, be it through major mergers and/or steady accretion, and that these ``assembly lines'' are stable over remarkably long time scales. Even one cluster without prominent large-scale filaments (MACS\,J0025.4$-$1222) exhibits an elongation of the outermost contours of the galaxy density near the main cluster that appears to point toward a satellite cluster, suggesting that an active filament may have been present in its direction at earlier times.

The most complex systems in terms of large-scale structure in this sample are MACS\,J0257.1--2335, MACS\,J0647.7+7015, and MACS\,J2214.9--1359. Each of these three features at least three collapsed cluster cores separated by several Mpc, as well as two nearly orthogonal axes of merging and infall along filaments. The large-scale morphology of the galaxy distribution in these fields thus strongly suggests that the respective MACS clusters occupy intersections of multiple filaments in the cosmic web. The detection of several filaments connecting satellite clusters to the main cluster provides strong observational support for the results of numerical simulations (Colberg et al.\ 2005; Colberg 2007) which suggest (a) that cluster pairs with very close separations ($<5$h$^{-1}$ Mpc) are always connected by filaments, and (b) that, statistically, the number of filaments increases with cluster mass.

A closer look at the prevalence of substructure in the cluster cores -- indicative of an ongoing merger -- and the strength of the filamentary extensions connecting these cores to their large-scale surroundings provides a fascinating snapshot of the wide variety in the dynamics and geometry of cluster growth. MACS\,J2214.9--1359, for instance, exhibits only subtle signs of disturbance in its core region (primarily a 90-degree rotation in the mild ellipticity of the galaxy-density isophotes from the very core to the outer regions) suggesting a state of advanced relaxation since the last major merger. On larger scales, however, faint filaments still trace the paths of past and future infall of matter. By stark contrast, the disturbed core of MACS\,J0647.7+7015 is clearly the scene of ongoing merging, and the strength of the filemantary connection to a series of satellite clusters to the southeast of the main cluster suggest that this phase of violent growth is to continue for many Gyr to come. Viewed as a whole, our twelve snapshots of the large-scale structure around the most massive galaxy clusters at $z\sim 0.55$ thus corroborate work by Colberg et al.\ (1999) whose numerical simulations show strong correlations between infall patterns and the large-scale structure surrounding massive clusters, with preferred directions for the assembly of clusters being established by tidal forces. Our observational results strongly suggest that, although multiple filaments are often present, one of them usually dominates in funneling matter onto the cluster occupying the respective vertex of the cosmic web.

Finally, a minority of our cluster targets exhibit noticeably elongated galaxy-density contours in their core regions (MACS\,J0744.8+3927, MACS\,J1149.5+2223, MACS\,J1423.8+2404), but little sign of substantial filaments on larger scales. Since the orientation of major filaments is random with respect to our line of sight it is possible that significant large-scale structure is present but not visible in projection. Indeed extensive spectroscopic follow-up of the galaxy population of these systems finds strong evidence for line-of-sight substructure in one of these three, MACS\,J1149.5+2223 (Smith et al.\ 2008, in preparation). At least one cluster in our sample can, however, be considered to be near virialization and safe from major mergers for some time, namely MACS\,J1423.8+2404 (see also the following section).

\subsection{X-ray Emission}

Although the small field of view, compared to SuprimeCam, of ACIS-I allows us to study only the X-ray emission from the main clusters shown in Fig. 2, fresh insights about the dynamical state of these clusters can be obtained by comparing the X-ray surface-brightness contours to the optical morphology of the main cluster (Fig. 3). In all cases we find the orientation of any elongation in the X-ray contours around the cluster core to be well aligned with the direction of infall as traced by the large-scale galaxy distribution. This is expected since the intra-cluster gas responds to the same gravitational forces as the galaxies and should thus, globally, show similar bulk motions.

There is a fundamental difference between gas and galaxies, however, in that the former is a strongly interacting viscous medium, whereas the latter are, on cluster scales, weakly or non-interacting point-like objects. As a consequence, gas and galaxies can show noticeable segregations in non-equilibrium situations like major cluster mergers. Indeed we find two systems for which the centroids of the gas and the galaxy distributions in the cluster cores are noticeably displaced from each other, namely MACS\,J0025.4--1222 and MACS\,J0717.5+3745, with the displacement corresponding to roughly 200--300 kpc in the plane of the sky. In either case the X-ray emission peaks between two well separated galaxy concentrations. A third, less obvious, example of the same phenomenon is MACS\,J0911.2+1746, a cluster consisting of two components of very different mass about 1 Mpc apart. A comparison of the galaxy and gas distributions around the less massive component shows the centroid of the X-ray emission to be significantly displaced from the peak in the galaxy surface density. The cause is, most likely, the same for all three systems mentioned, namely an ongoing merger in which the galaxy populations of the two merging clusters have already passed through each other with little interaction, whereas the collision of the dense gas in the cluster cores creates shocks through which kinetic energy from the bulk motion of the gas is partly converted into thermal energy and partly redirected into transverse velocity components.

Similar cases of a clear segregation between gas and galaxies have been observed in other clusters and have helped greatly to improve our understanding of cluster mergers. A particularly spectacular example is the ``bullet cluster'' 1E0657--558 (Markevitch et al.\ 2002) in which the fortuitous alignment of the merger axis with the plane of the sky enabled remarkably detailed and quantitative studies of the dynamics of the merging process. It remains, so far, the only system to allow direct constraints to be placed on the dark matter self-interaction cross section (Markevitch et al.\ 2004, Clowe et al.\ 2006). The potential implications of significant offsets between the centroids of the gas, galaxy, and dark-matter distribution in clusters are thus far-reaching, and further follow-up studies, in particular a gravitational-lensing analysis of the total-mass distribution, of MACS\,J0025.4--1222, MACS\,J0717.5+3745, and MACS\,J0911.2+1746 may offer a rare opportunity to test the results obtained for 1E0657--558 on the nature and fundamental properties of dark matter.

Two of our 12 MACS clusters (MACS\,J0744.8+3927 and MACS\,J1423.8+2404) deserve special mention as the only ones showing obvious cooling cores in our X-ray images. MACS\,J1423.8+2404 is also the sole cluster that shows neither obvious large-scale structure nor significant substructure in the core region. Although, again, structure along the line of sight cannot be ruled out conclusively, the system's optical and X-ray appearance suggests that it is close to virialization (see also Ebeling et al.\ 2007). While a cooling core is also clearly present in MACS\,J0744.8+3927, the large-scale environment of this system and thus its overall assembly history and future are, at present, difficult to assess owing to the limited depth of the existing SuprimeCam data for this most distant cluster in our sample.

\section{Summary}

We use the projected surface density of galaxies near the cluster red sequence to trace the large-scale environment around all twelve MACS clusters at $z > 0.5$ from the sample of Ebeling et al.\ (2007). Since the galaxy-color criterion applied selects primarily early-type galaxies dominated by old stellar populations this approach is well suited to map the galaxy distribution in regions of moderate to high density (cluster cores, satellite clusters, and the densest regions of the filamentary structures connecting them), but will be less efficient in unveiling low-density structures such as galaxy sheets or the environment customarily referred to as the ``infall region'', i.e.,\ the region around the cluster--field interface. Our study uses wide-field imaging data obtained with the SuprimeCam imager on the Subaru 8m telesope on Mauna Kea and probes scales of typically 10 Mpc at the median redshift of $z=0.55$ of our cluster targets. By correcting the distribution of observed galaxy magnitudes to a common reference redshift (again $z=0.55$) and applying a global magnitude cut we ensure that all systems are sampled uniformly to 0.03$L_R^\star$, with the exception of MACS\,J0744.8+3927 for which our existing data become incomplete at 0.2$L_R^\star$.

We find the vast majority of these most massive clusters at $z\sim 0.55$ to be embedded in a highly complex large-scale environment characterized by satellite groups/clusters and extensive filaments. In all cases the alignment of any substructure and/or ellipticity of the galaxy distribution in the cluster cores with the filaments found on much larger scales strongly suggests that the ongoing growth of these systems occurs through merging and infall of matter along a single preferred axis that is stable over many Gyr. Several clusters are found to reside at the intersection of multiple distinct axes of infall which appear to have actively contributed to the assembly of the cluster in the past, presumably alternating in dominance. The observed large-scale morphologies and their implications for the dynamics of cluster formation and evolution thus provide strong observational support for the results of numerical work that finds the growth of the most massive galaxy clusters to occur at the densest vertices of the cosmic web, tightly linked to filaments funneling matter onto
the central cluster. Since this work, by design, is biased toward detecting the denser regions of the environments studied, the extensive filaments found should be well suited for observational attempts to directly detect and study the properties of the ``Warm-Hot Intergalactic Medium'' (WHIM) hypothesized to contain a significant fraction of the baryons in the universe.

Chandra observations of all clusters in our sample allow a comparison of the distribution of galaxies with that of the gas in the central cluster regions. We find the morphology of gas and galaxies to be similar on large scales, consistent with the expectation that in near-equilibrium conditions all cluster components participate similarly in gravitation-induced bulk motions. We, however, find significant offsets between the centroids of the two distributions in the cases of MACS\,J0025.4--1222, MACS\,J0717.5+3745, and MACS\,J0911.2+1746. All of these systems are thus promising targets for in-depth follow-up studies of the distributions of gas, galaxies, and dark matter in merging clusters that hold great promise for advancing our understanding not just of the physics and dynamics of cluster mergers but specifically also of the nature and properties of dark matter (Brada\v{c} et al.\, in preparation).

Finally, the wide range of environments discovered here confirms MACS clusters as ideal targets for detailed studies of the impact of environment on galaxy evolution. First results on the distribution and origin of E+A galaxies from an extensive spectroscopic follow-up campaign of galaxies in the field of MACS\,J0717.5+3745 underline the significant advantage of being able to sample the full range of environments, from the low-density field to the extremely dense cluster cores, in a single observational target (Ma et al.\ 2008).

We would like to thank Leif Wilden and Glenn Morris of Stanford University for assistance with the reduction of some of the optical data used in this study. HE gratefully acknowledges financial support from NASA LTSA grant NAG 5-8253 and SAO grants GO2-3168X, GO3-4168X, and GO3-4164X.

\clearpage
\begin{figure*}
\includegraphics[width=3in]{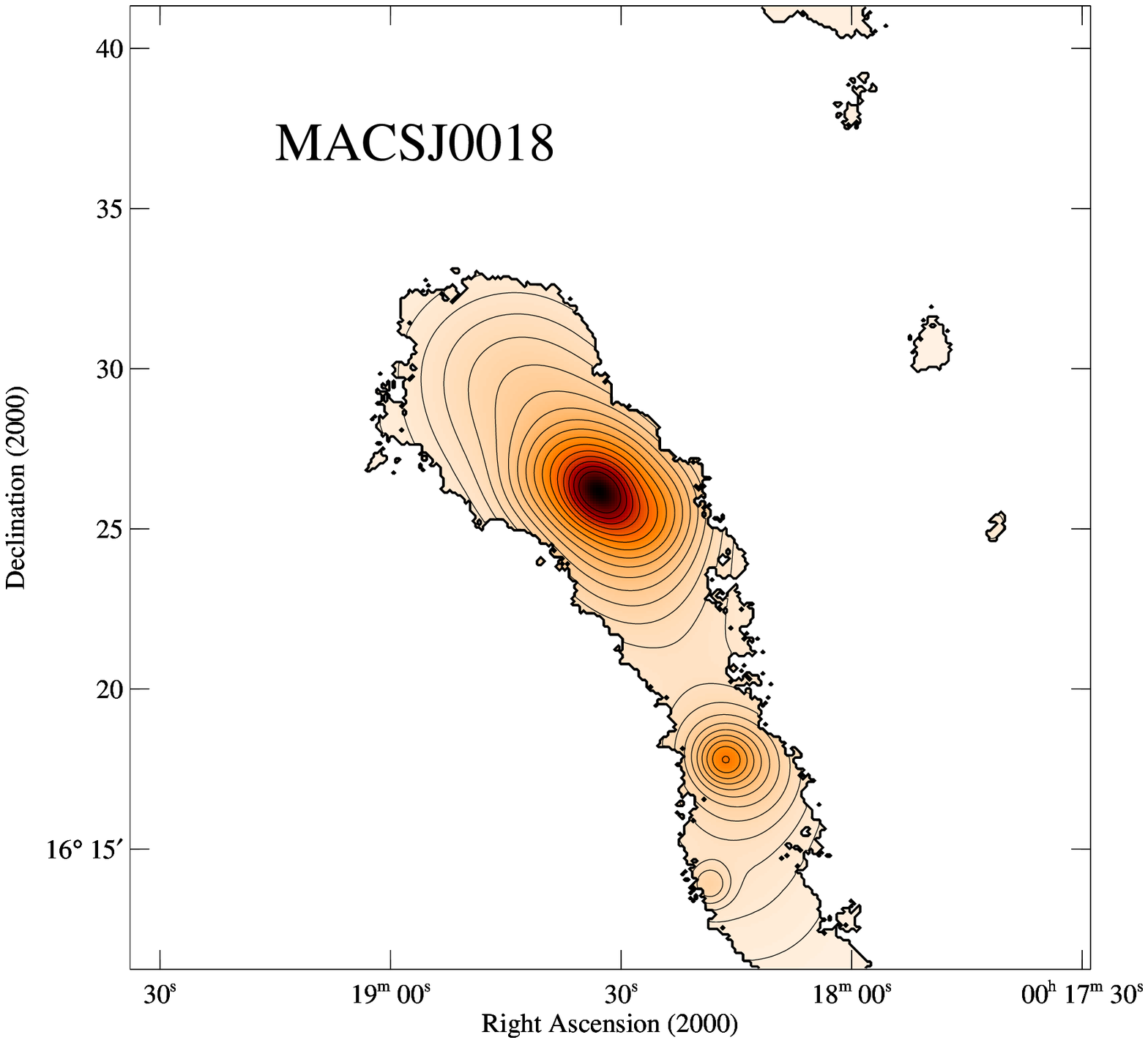}
\includegraphics[width=3in]{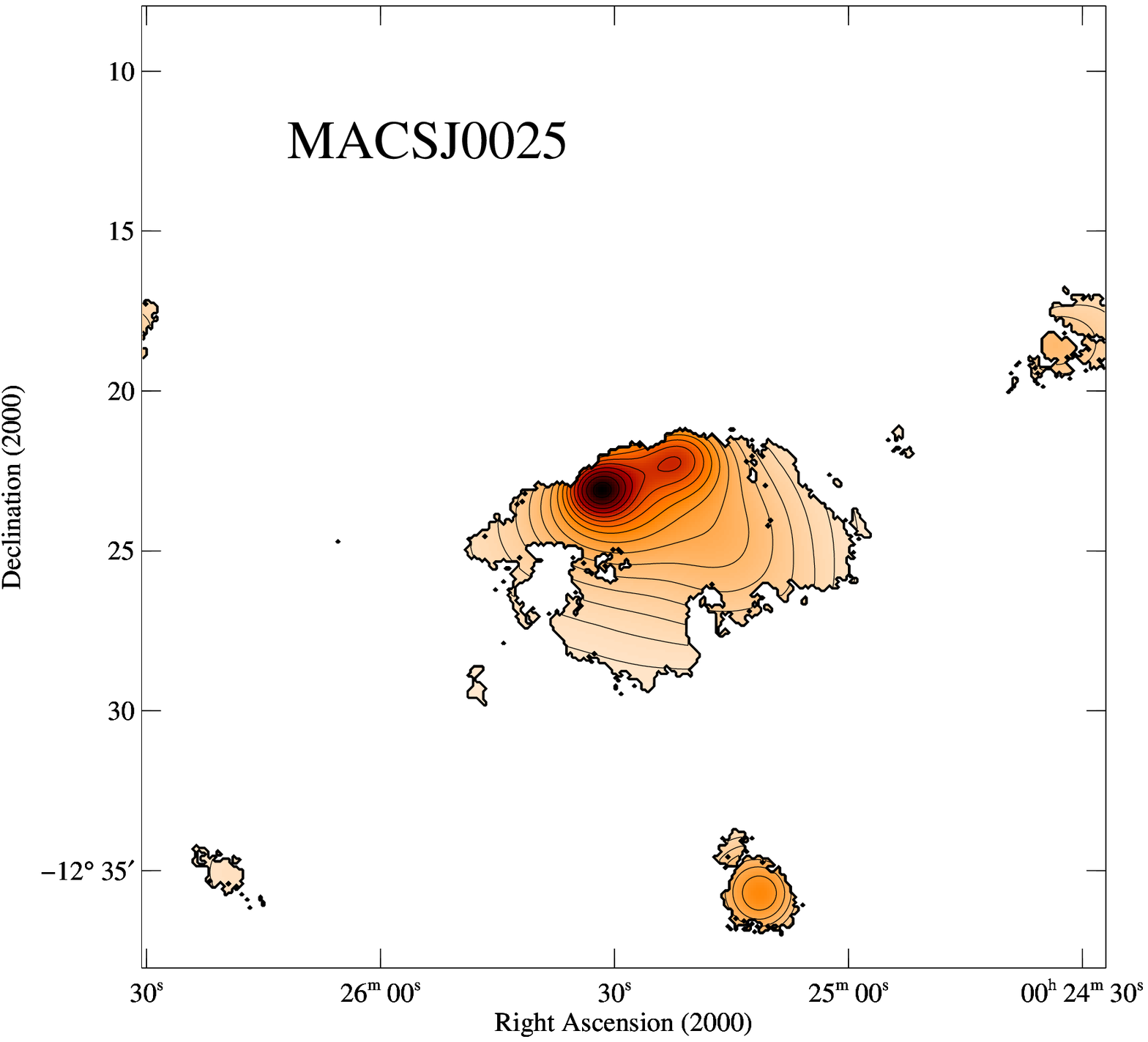} \\
\includegraphics[width=3in]{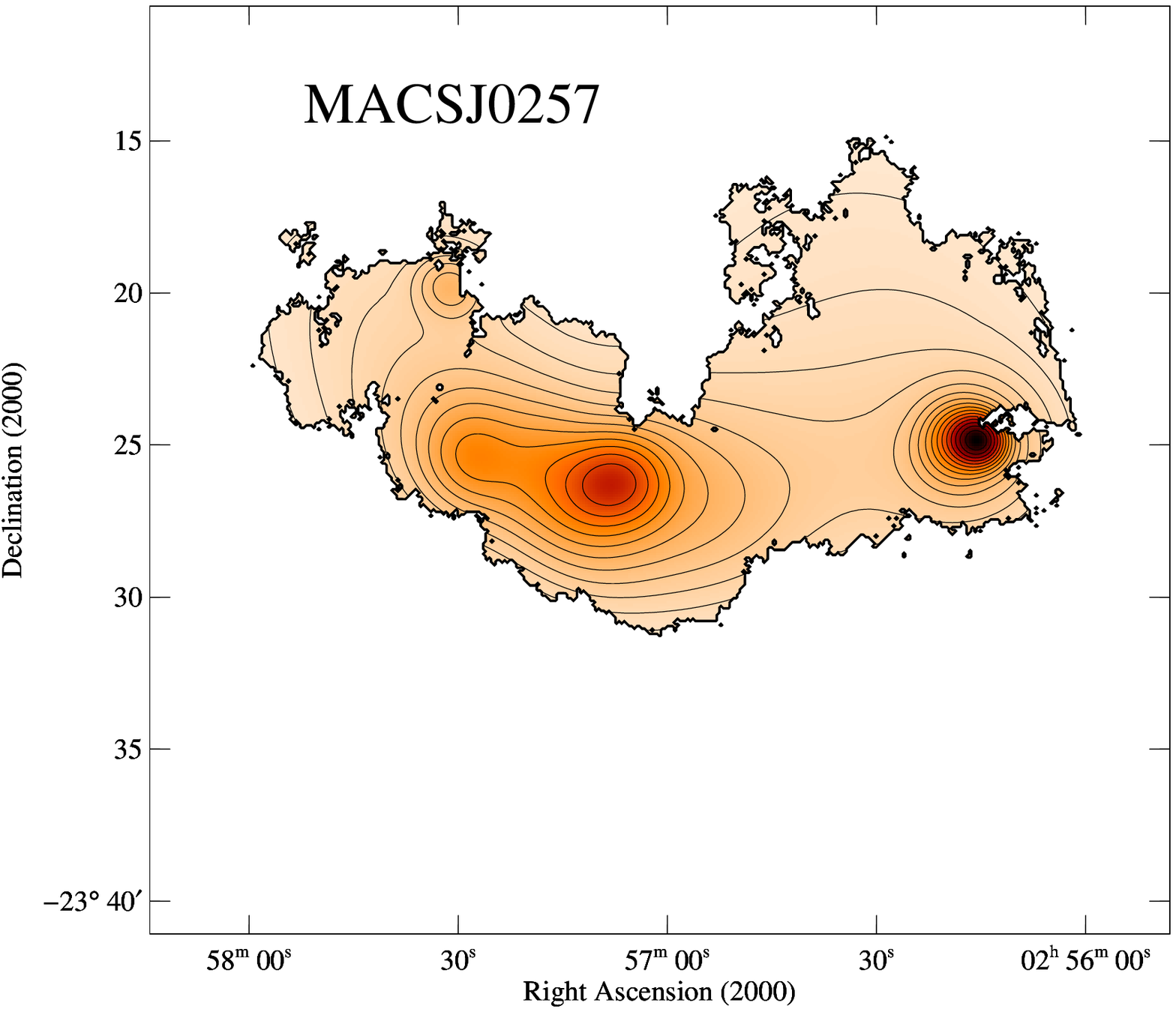}
\includegraphics[width=3in]{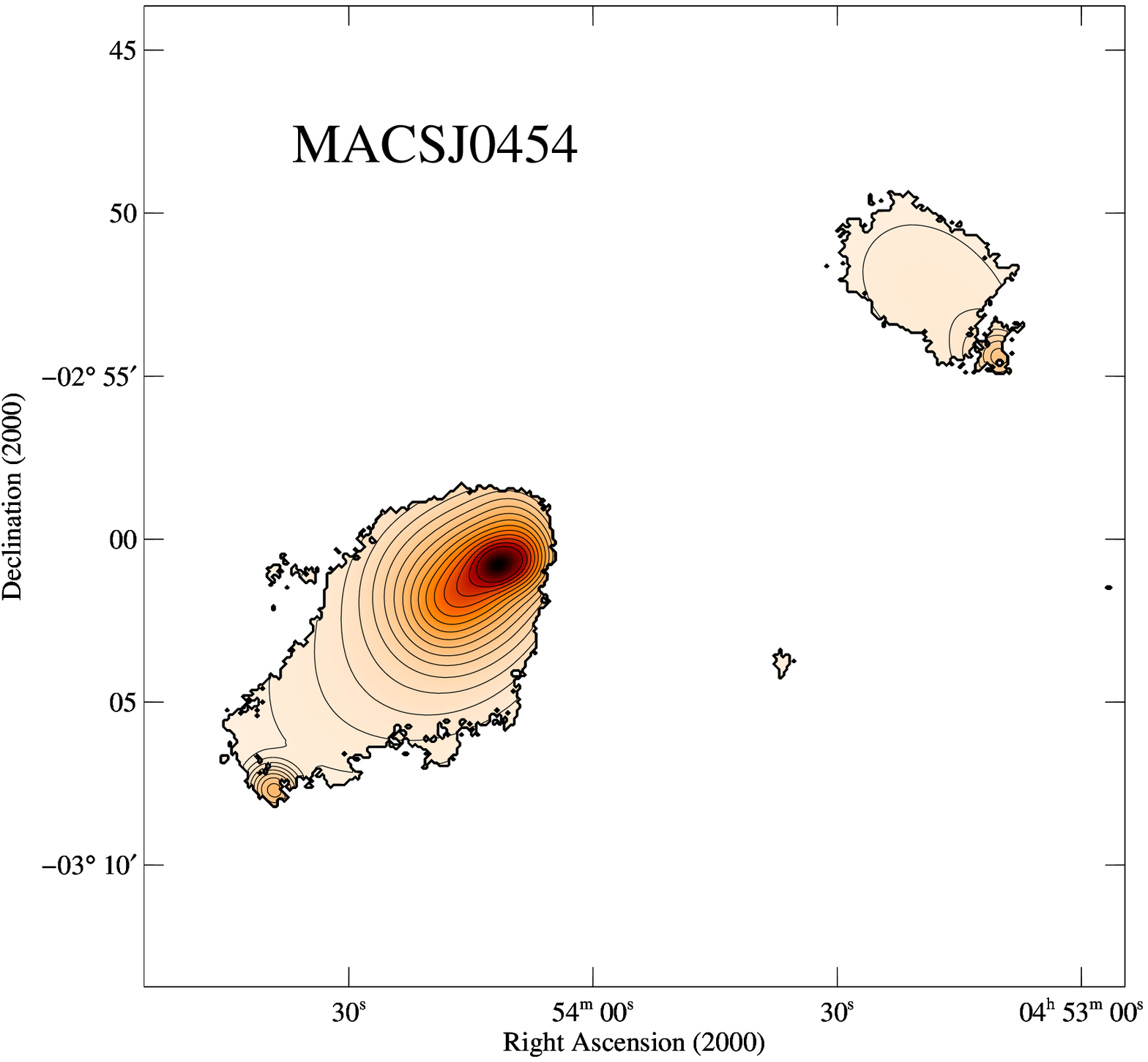} \\
\includegraphics[width=3in]{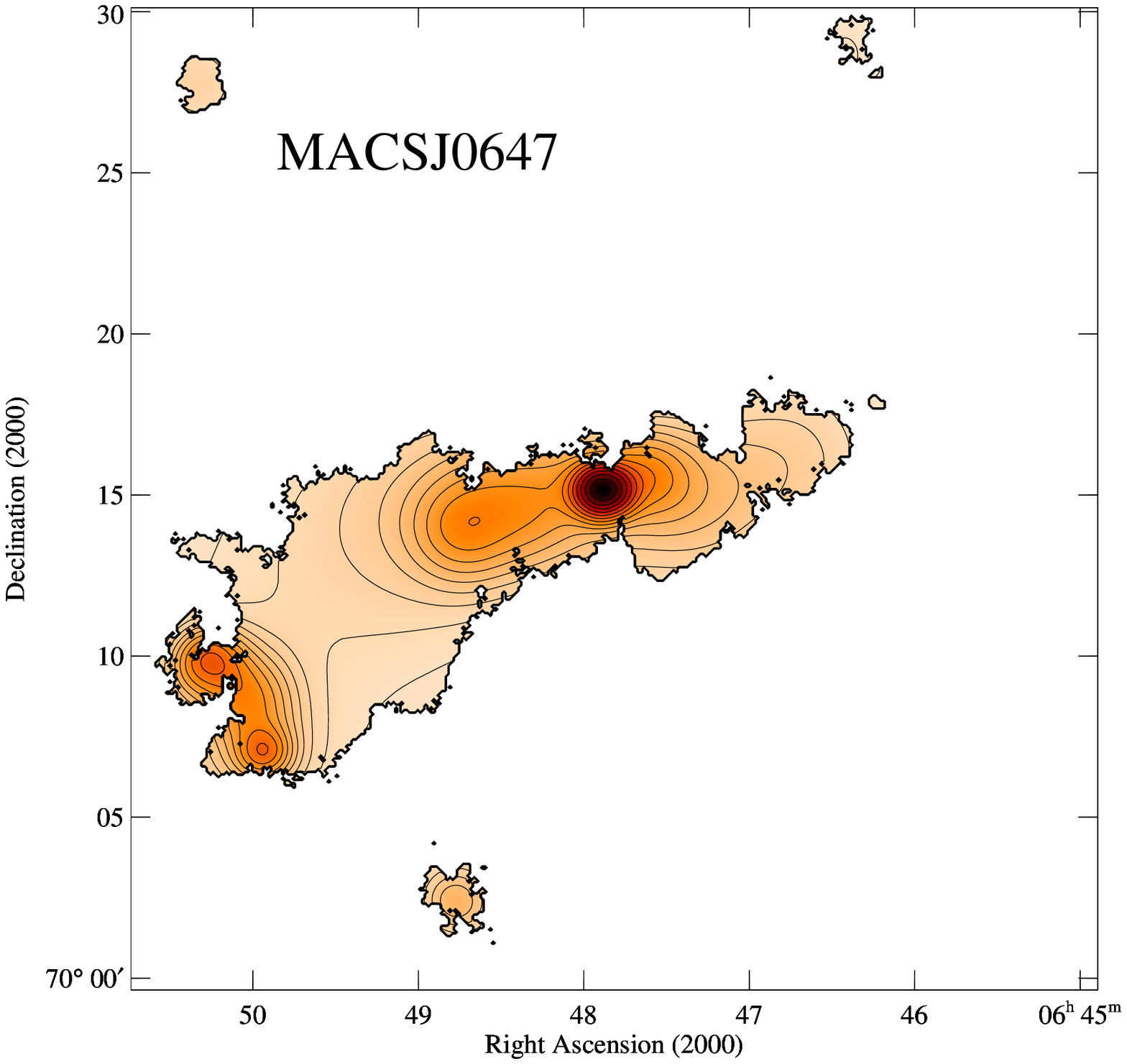}
\includegraphics[width=3in]{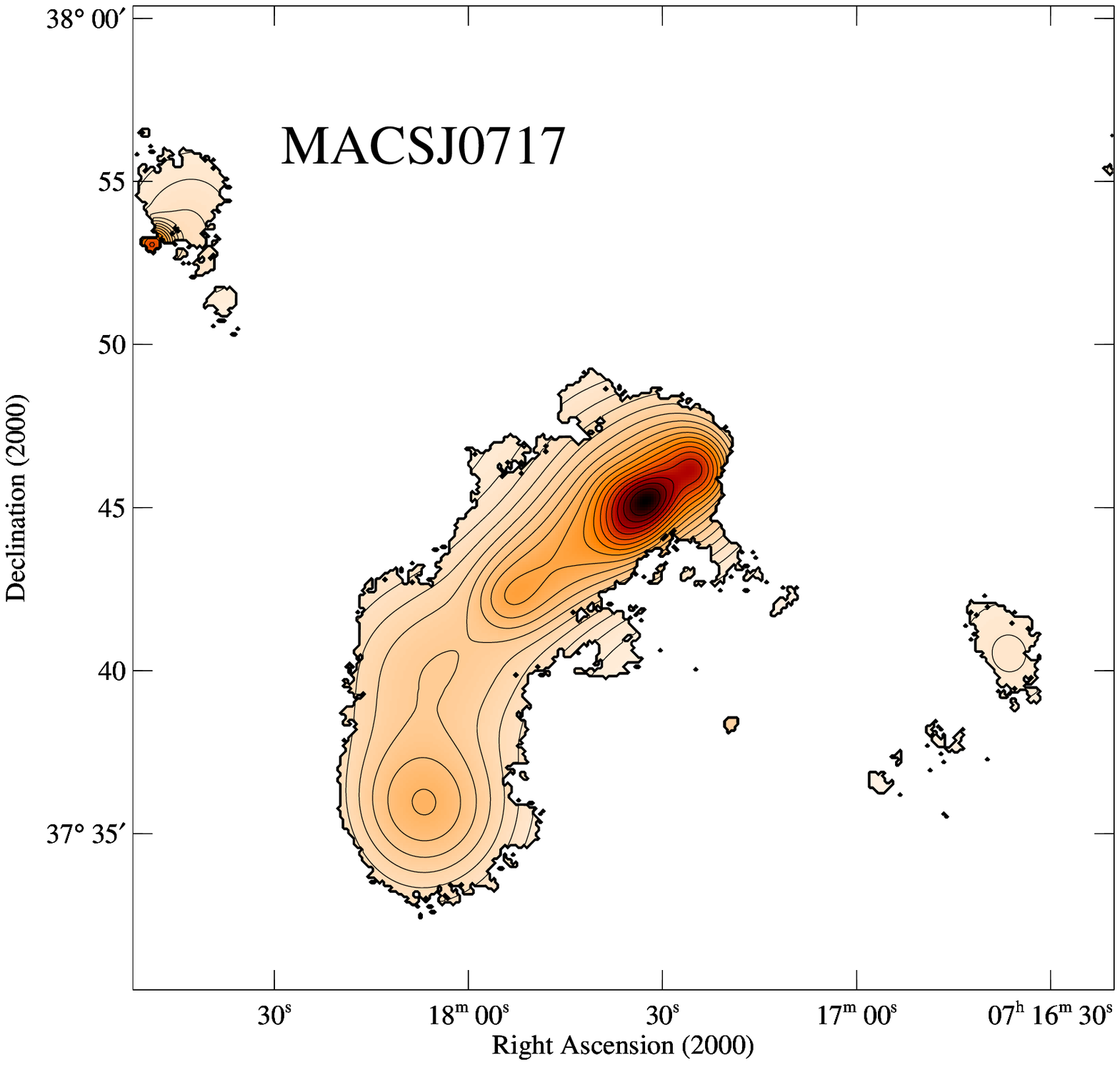} \\
\caption{Contours of the the adaptively smoothed projected surface-density distribution of galaxies on the cluster red sequence.  The contours are logarithmically spaced by $20\%$. Features within the bold line are significant at the 3$\sigma$ confidence level.}
\end{figure*}

\clearpage
\begin{figure*}
\noindent
\includegraphics[width=3in]{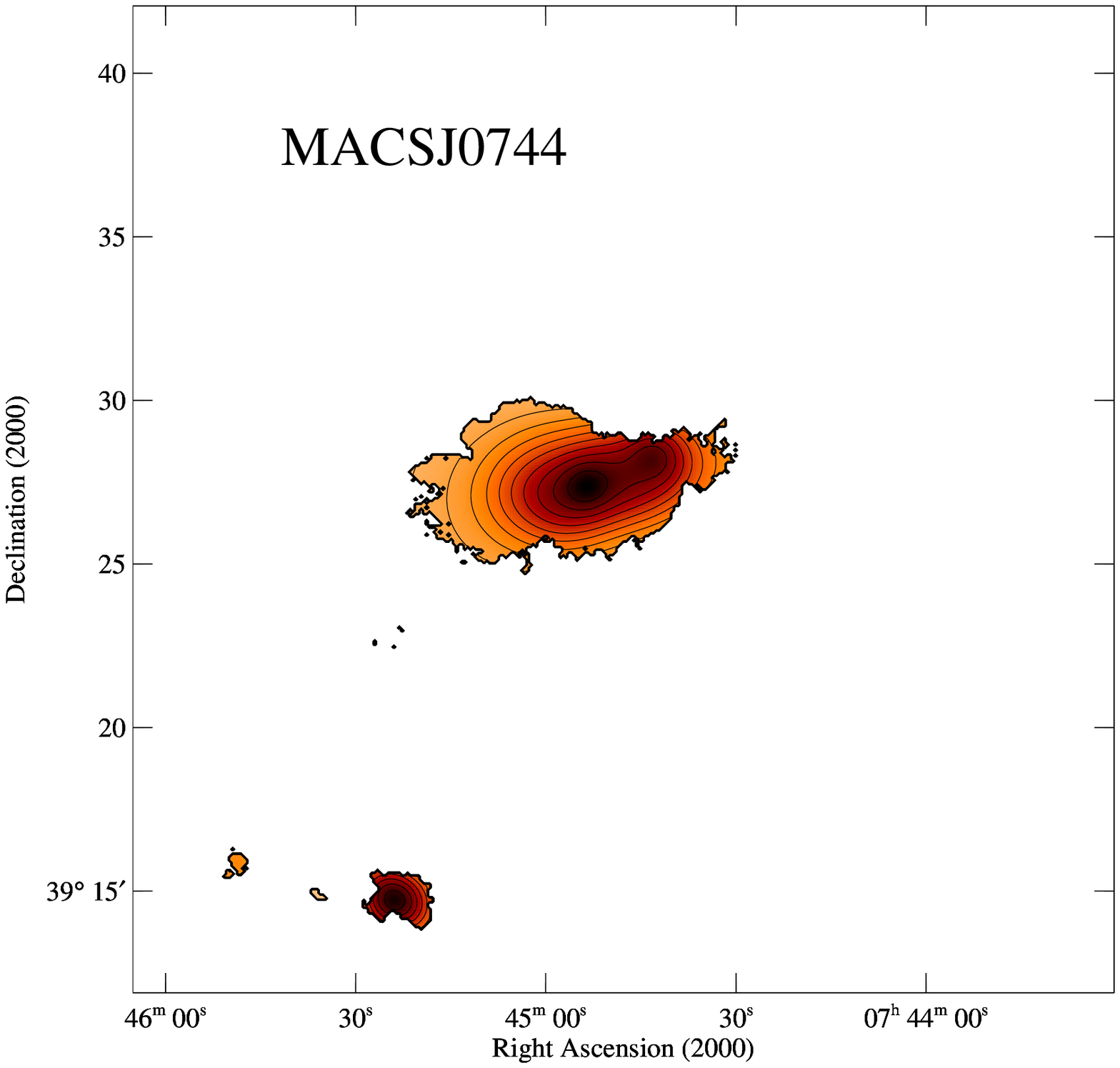}
\includegraphics[width=3in]{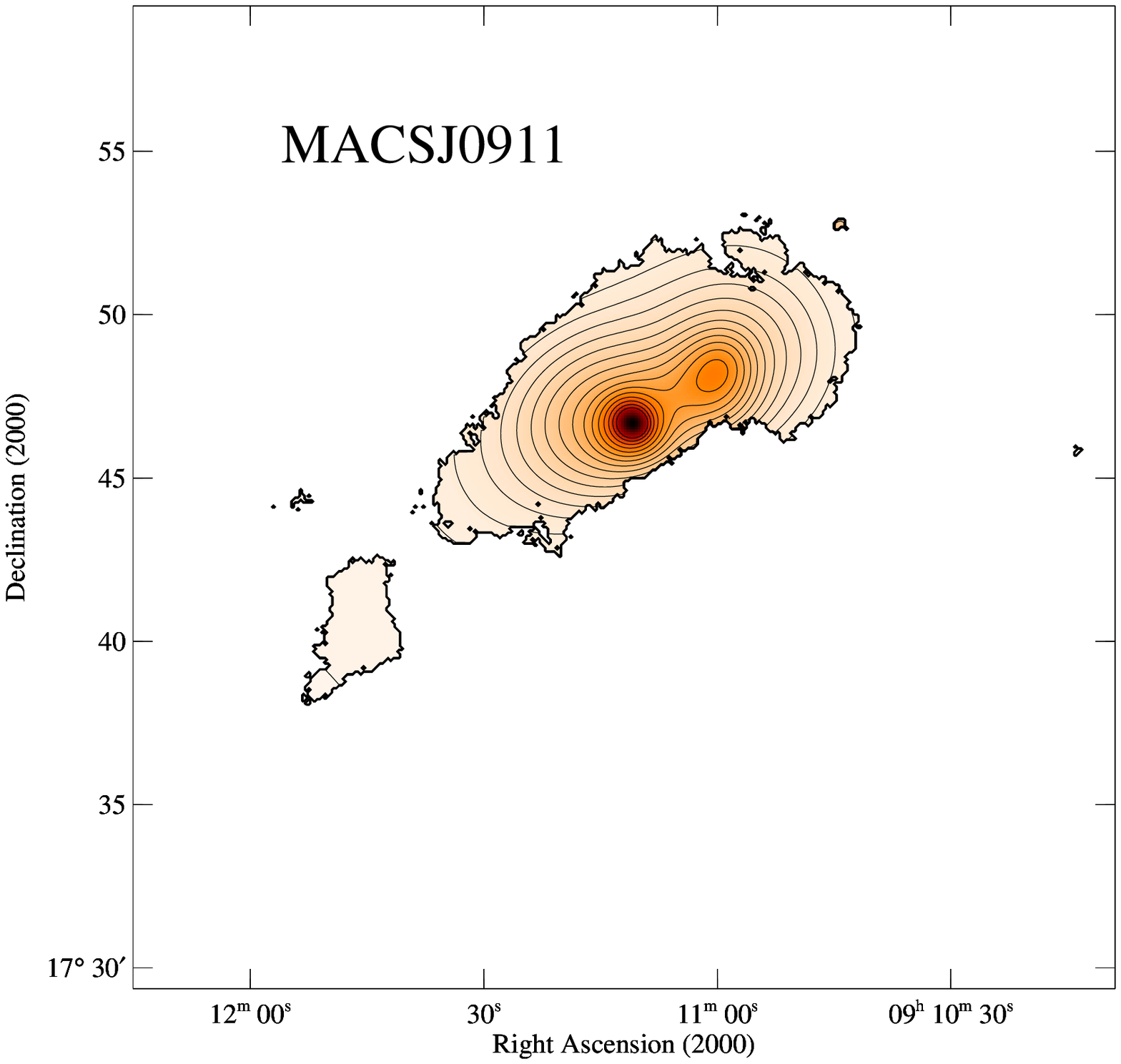} \\
\includegraphics[width=3in]{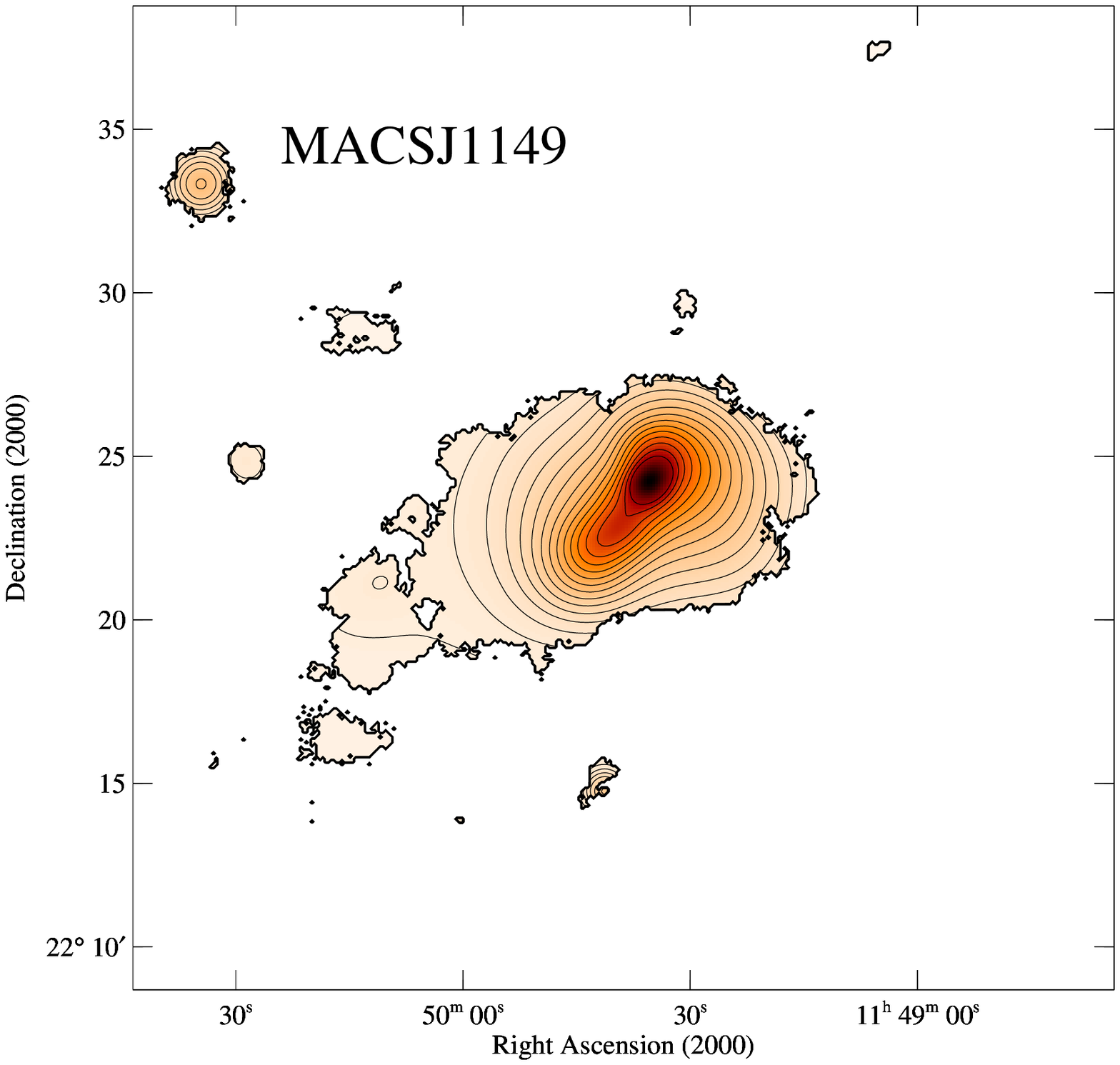}
\includegraphics[width=3in]{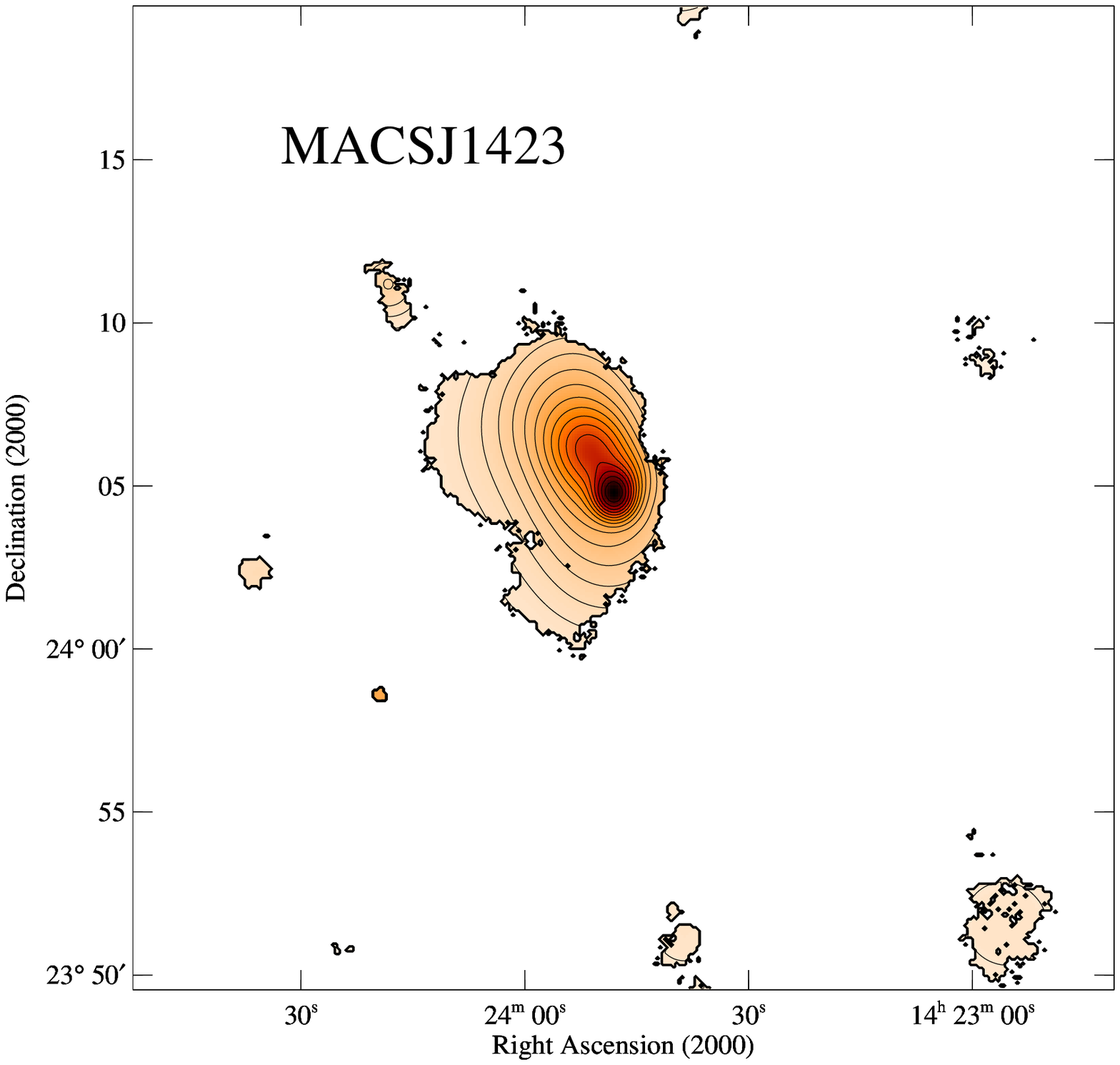} \\
\includegraphics[width=3in]{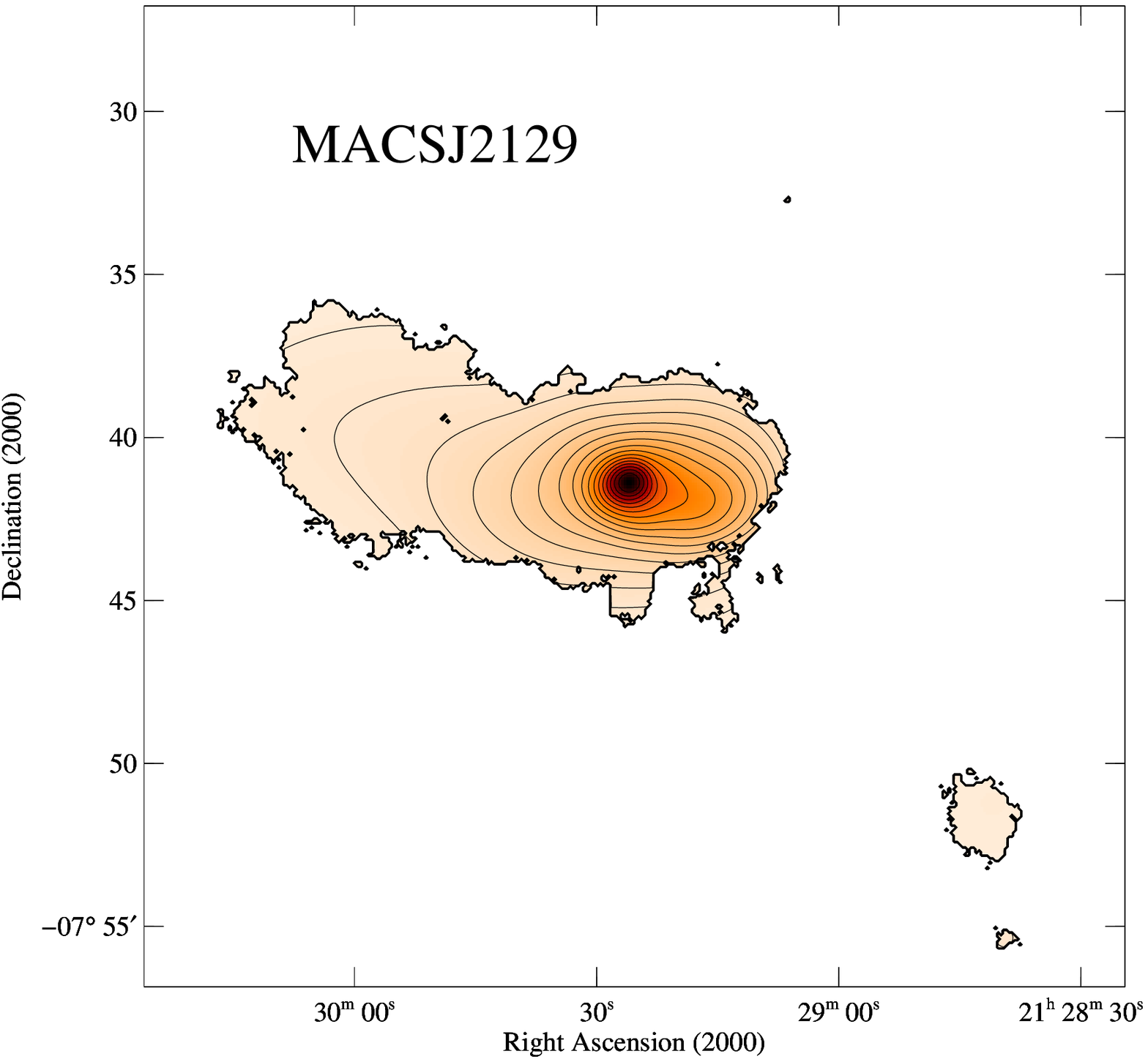}
\includegraphics[width=3in]{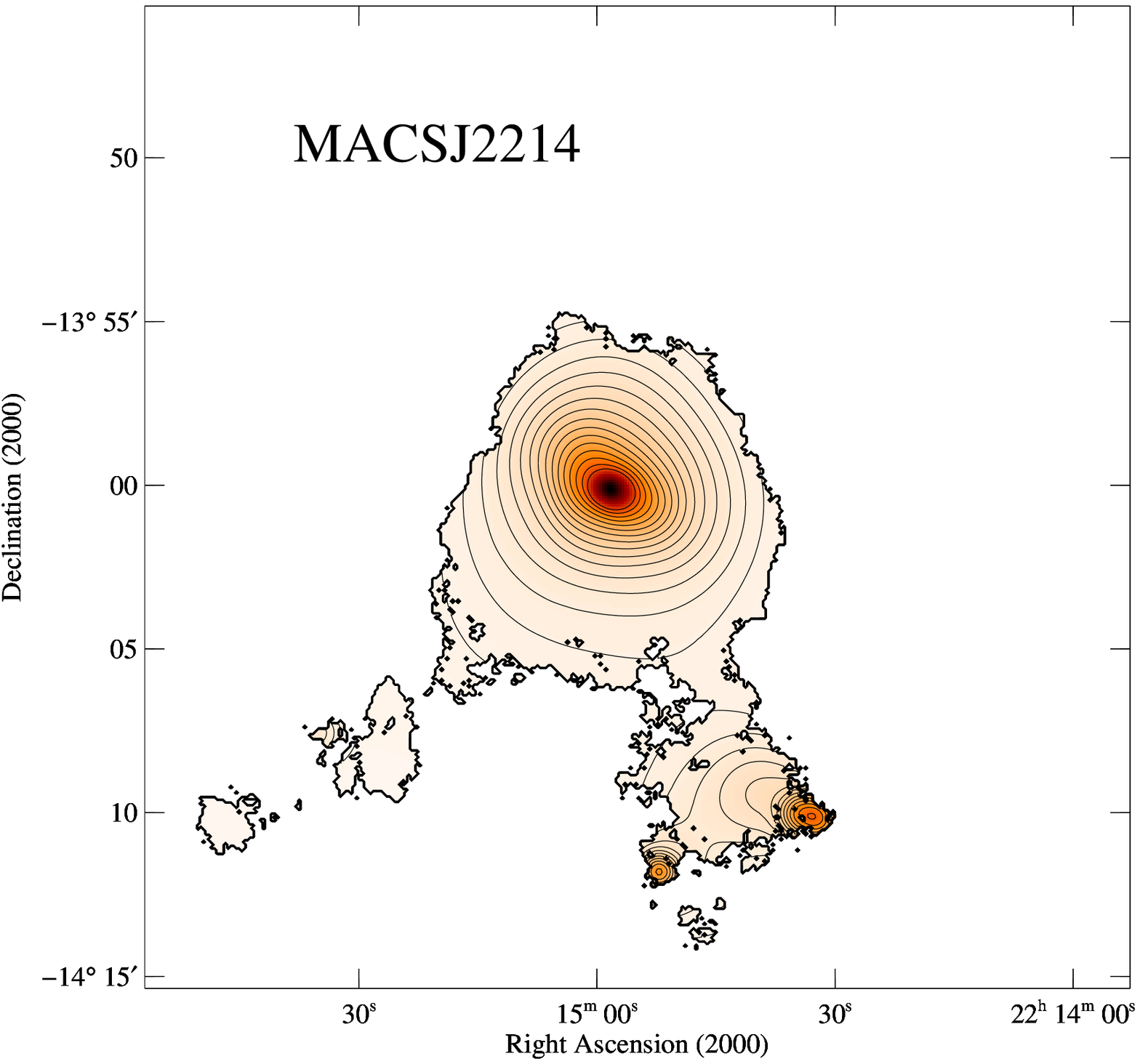} \\
\contcaption{}
\end{figure*}

\clearpage
\begin{figure*}
\includegraphics[width=3in]{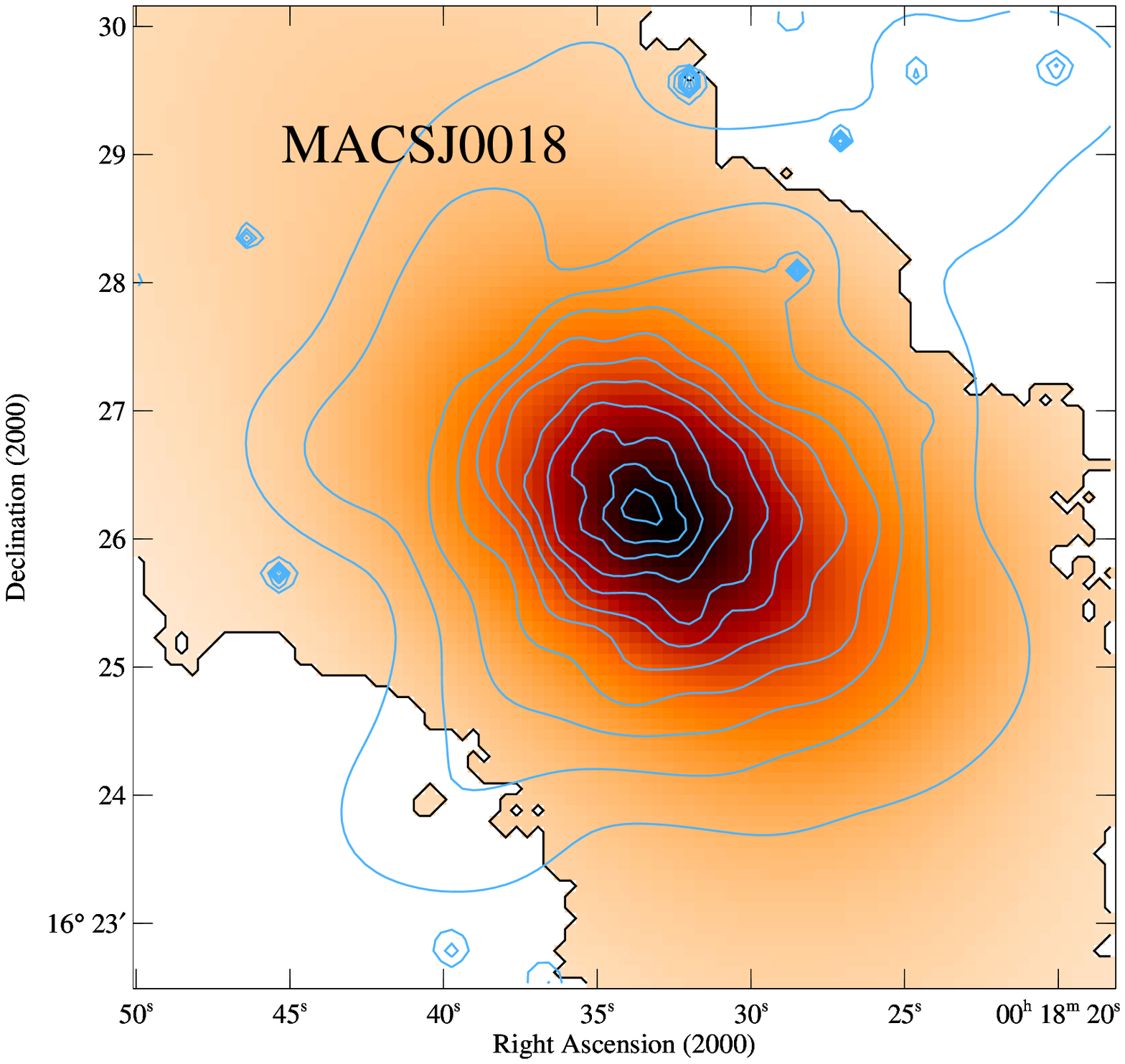}
\includegraphics[width=3in]{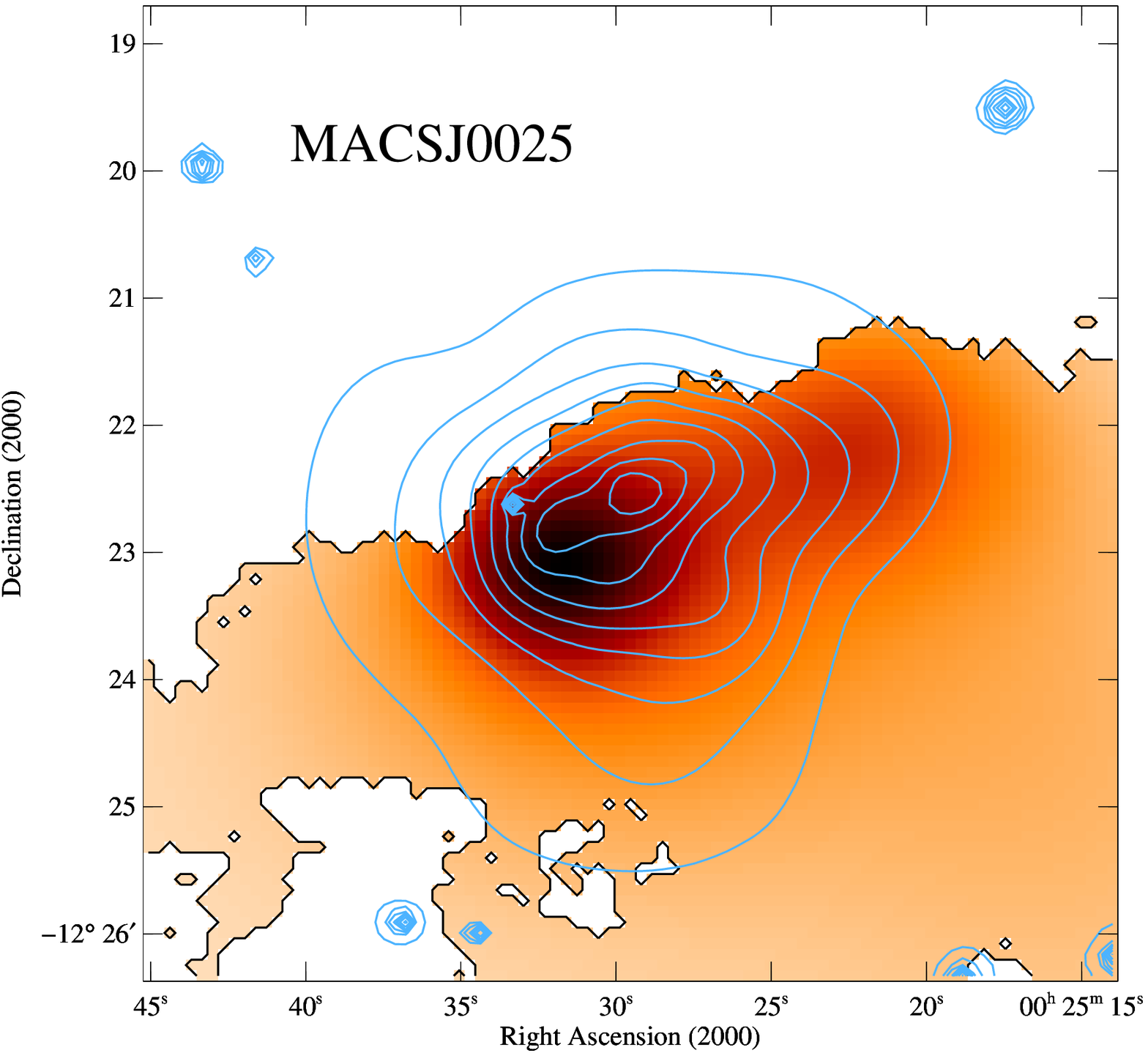} \\
\includegraphics[width=3in]{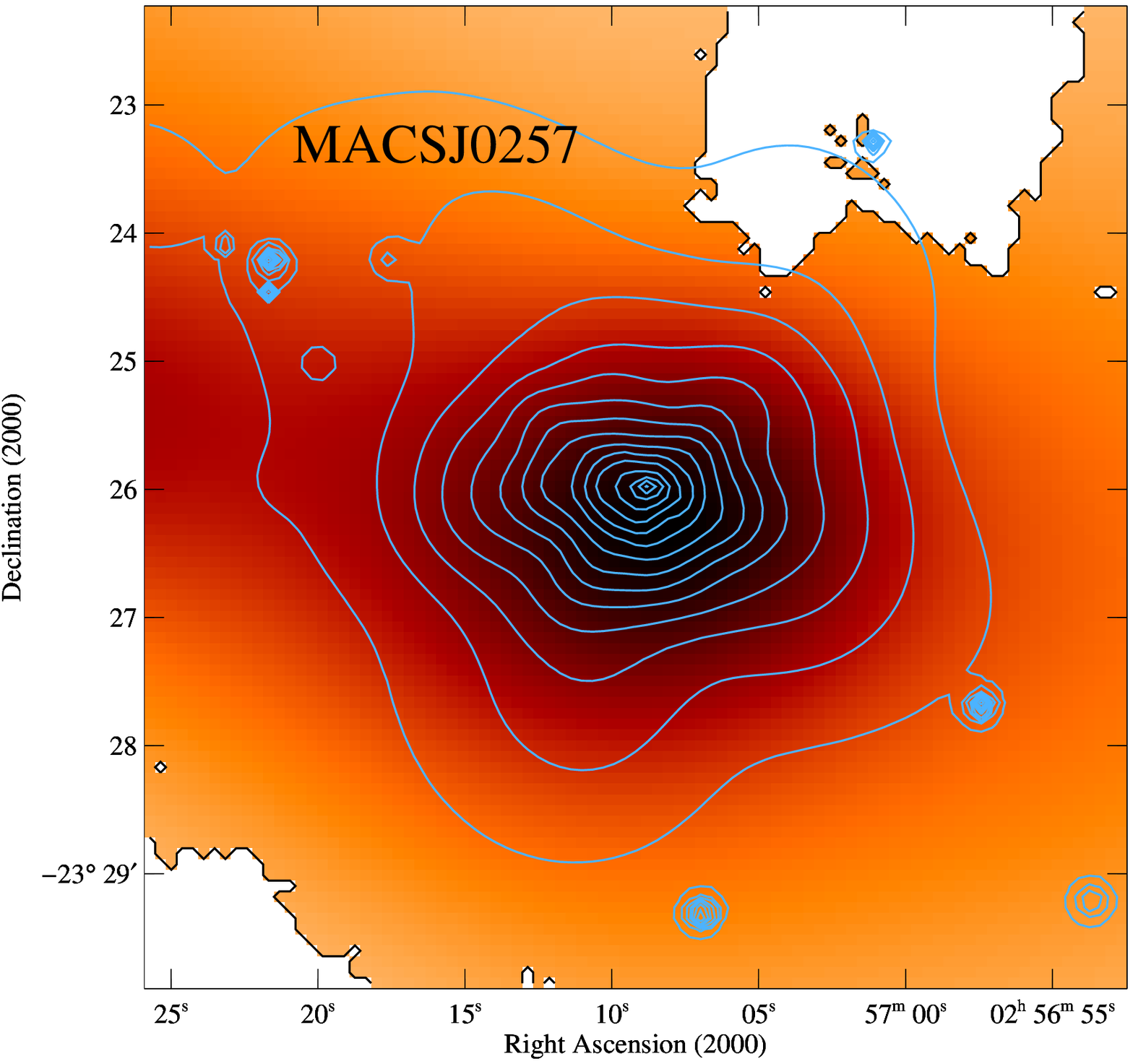}
\includegraphics[width=3in]{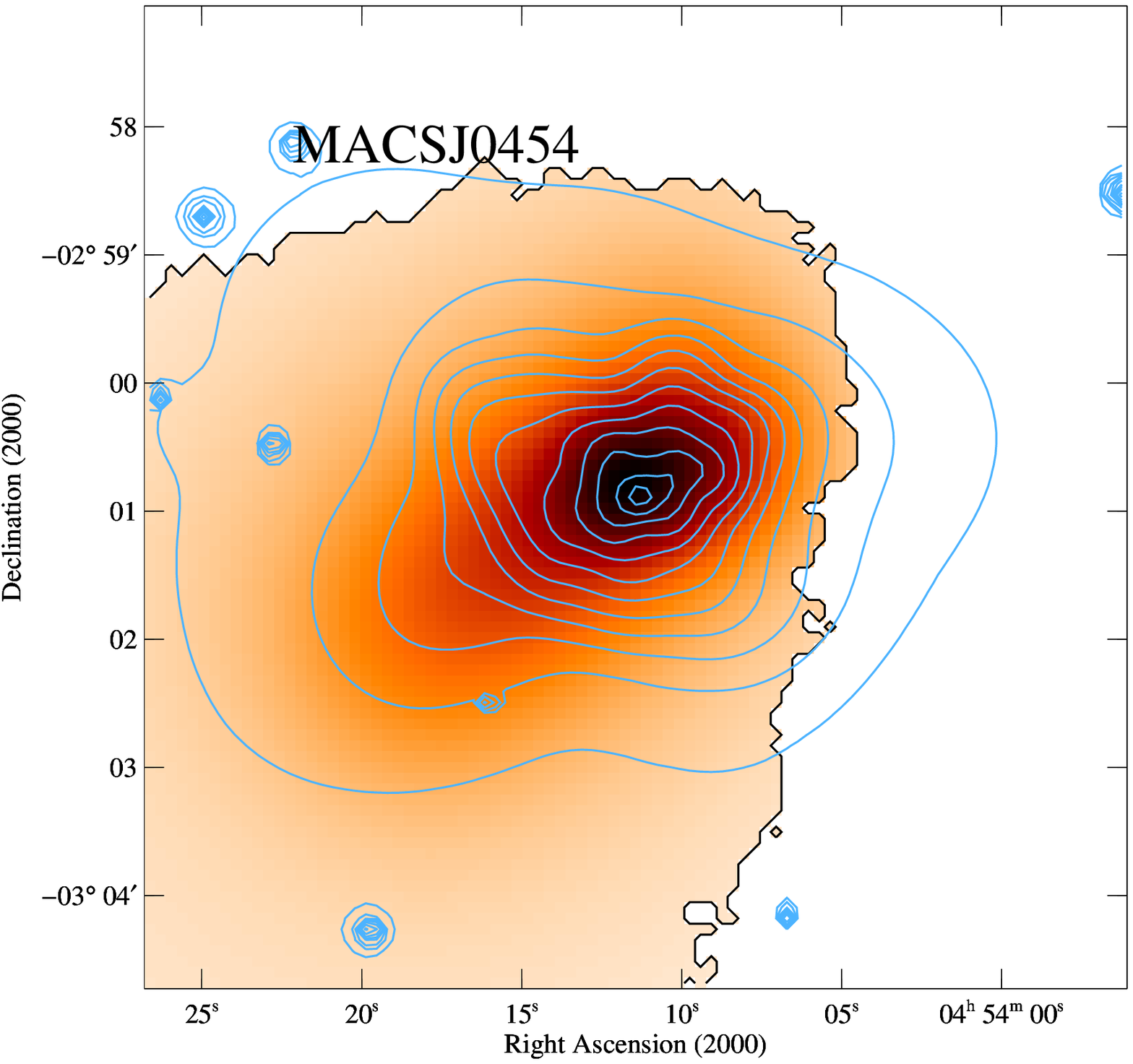} \\
\includegraphics[width=3in]{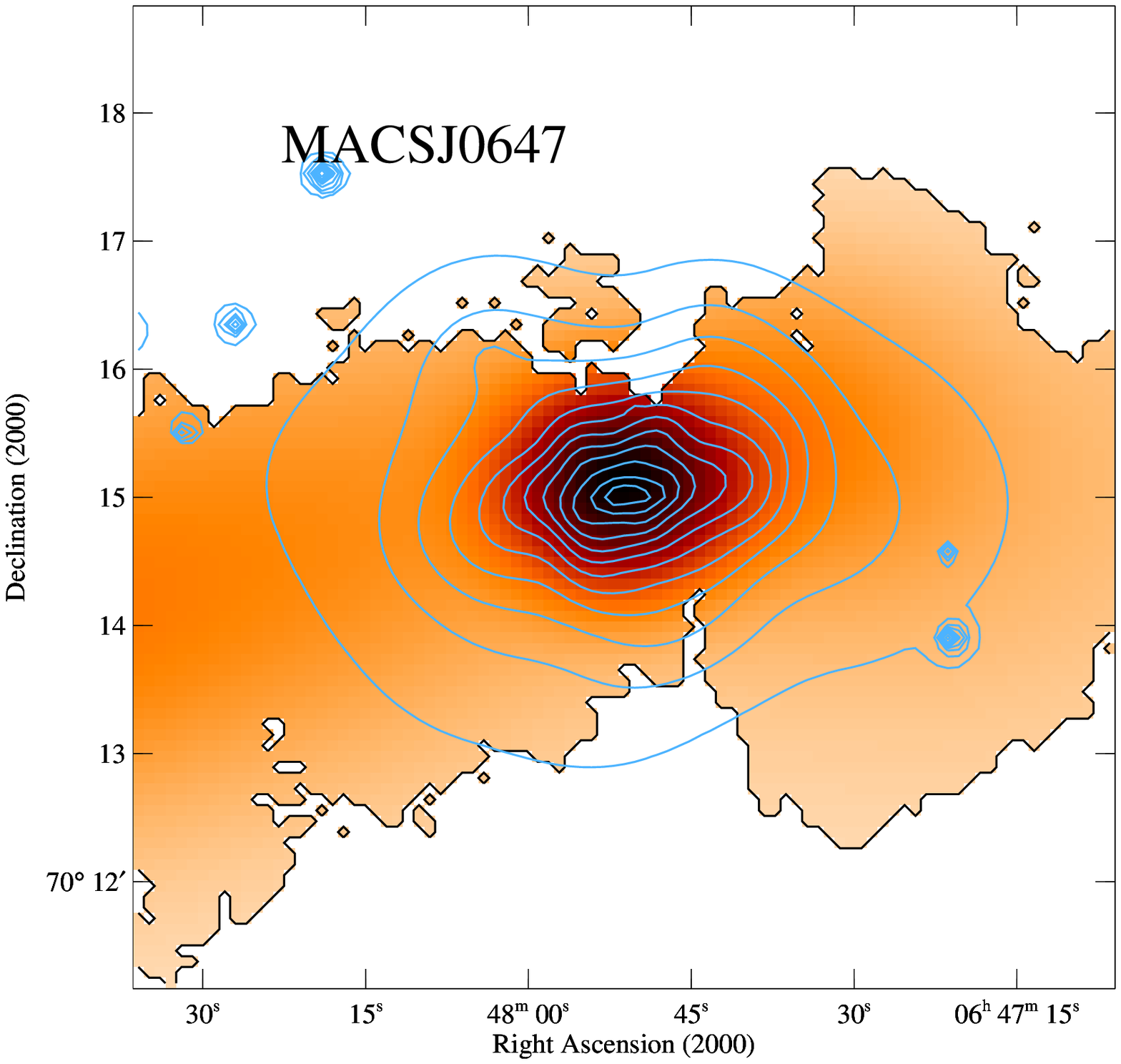}
\includegraphics[width=3in]{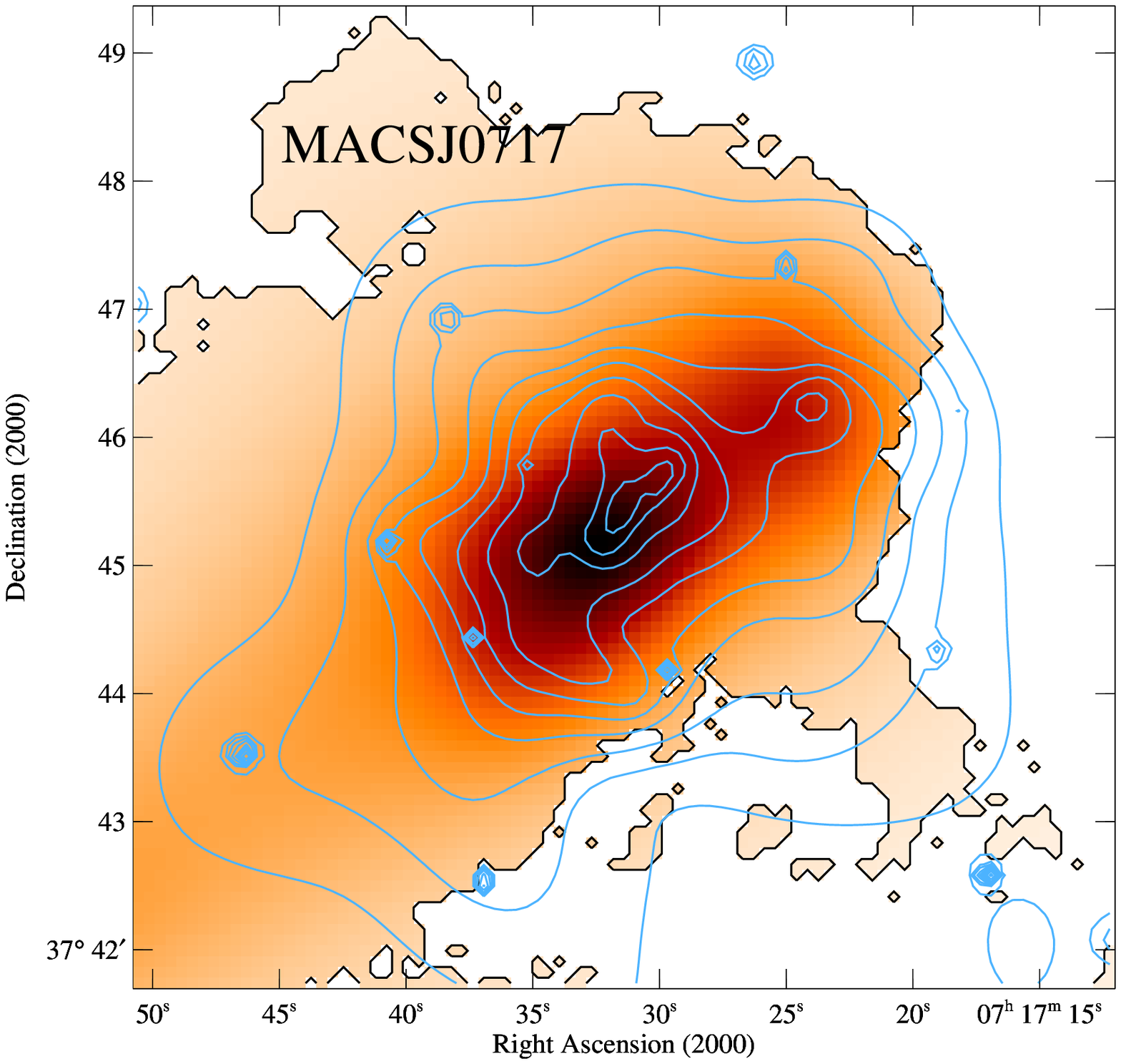} \\
\caption{Scaled representation of the same data as in Fig. 2 but showing only the core regions of each cluster. Overlaid in blue are the contours, logarithmically spaced by $20\%$, of the adaptively smoothed X-ray emission from the diffuse intra-cluster medium as observed with Chandra/ACIS-I in the 0.5--7 keV band.}
\end{figure*}

\clearpage
\begin{figure*}
\includegraphics[width=3in]{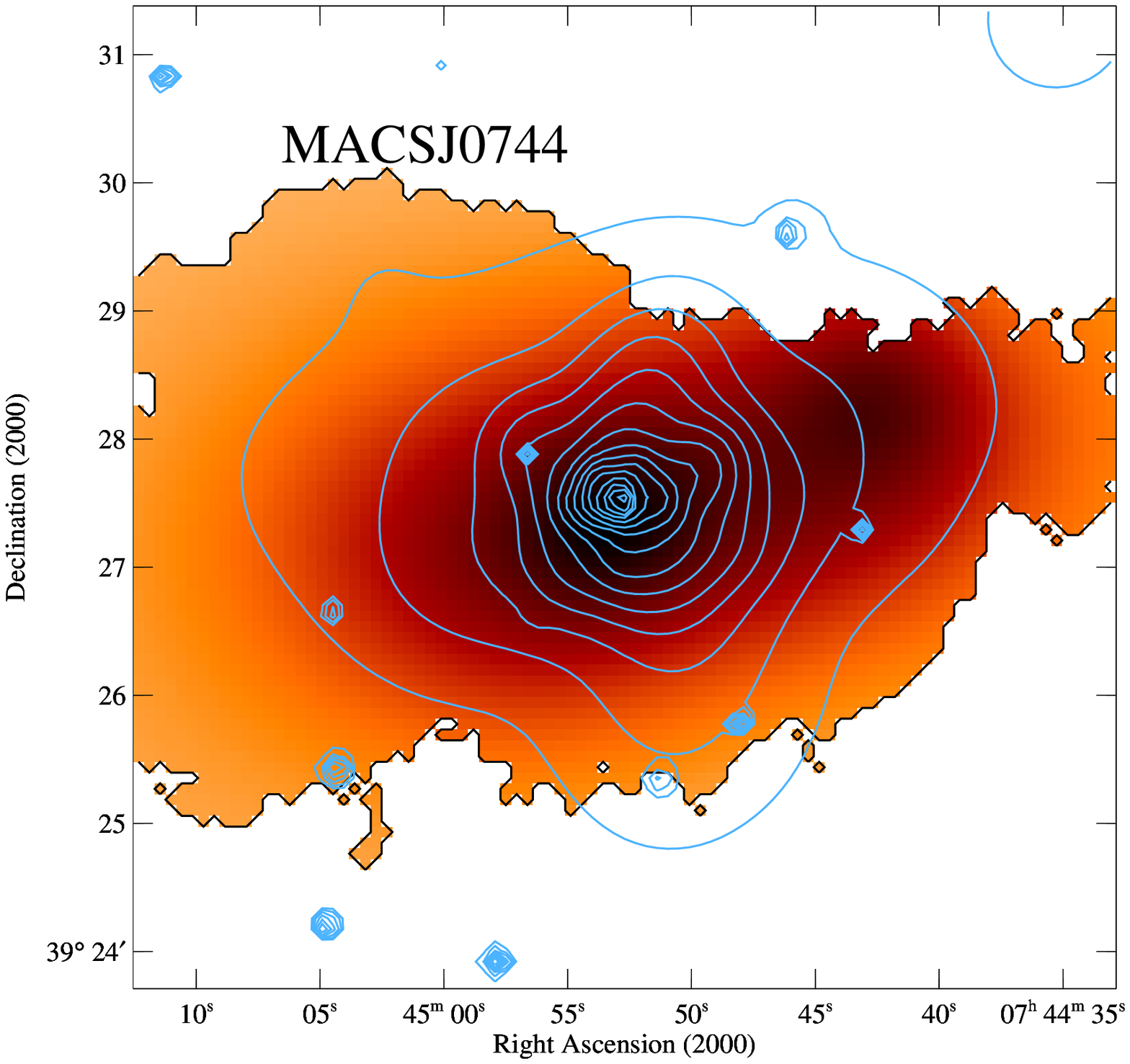}
\includegraphics[width=3in]{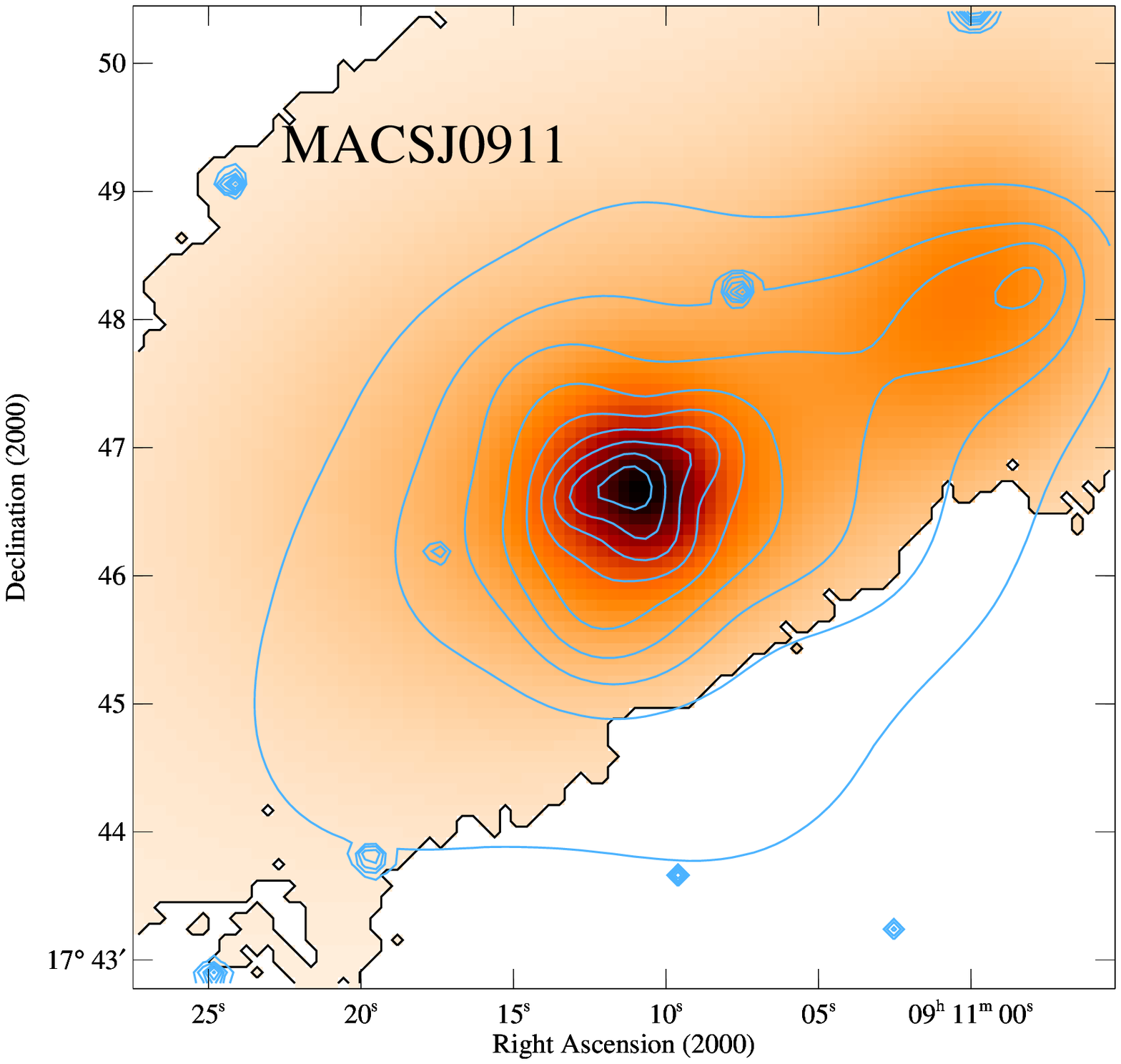} \\
\includegraphics[width=3in]{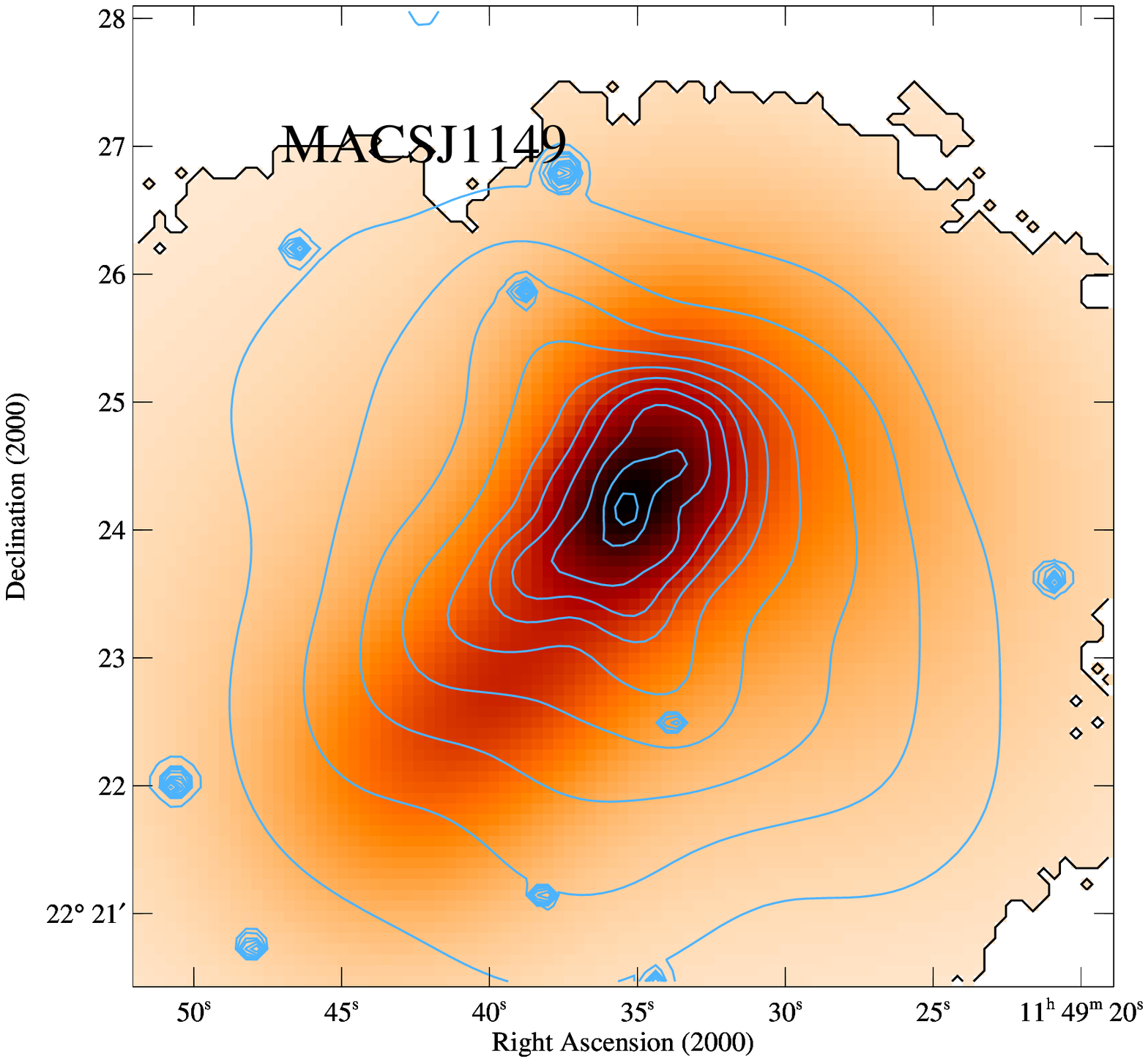}
\includegraphics[width=3in]{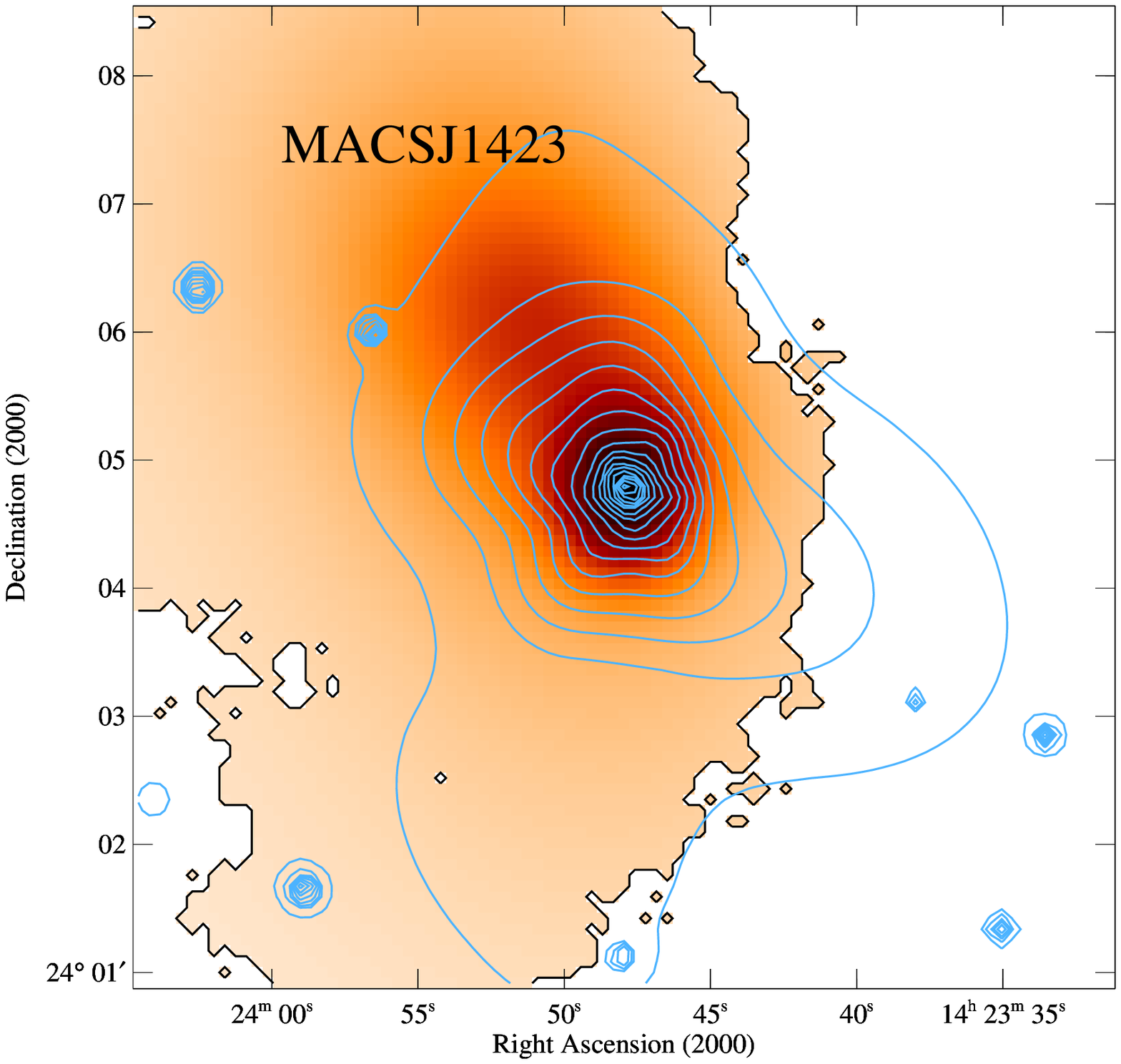} \\
\includegraphics[width=3in]{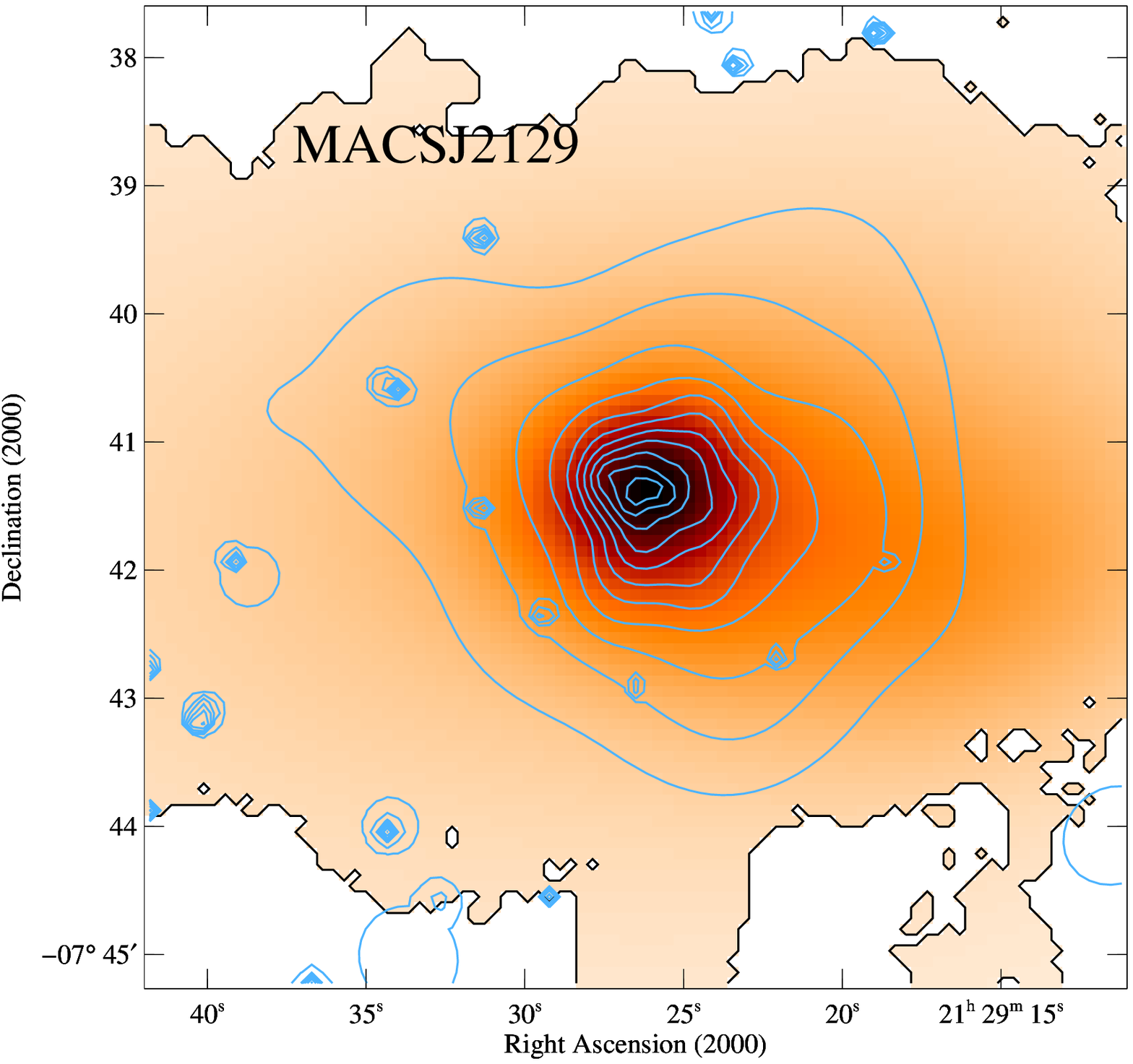}
\includegraphics[width=3in]{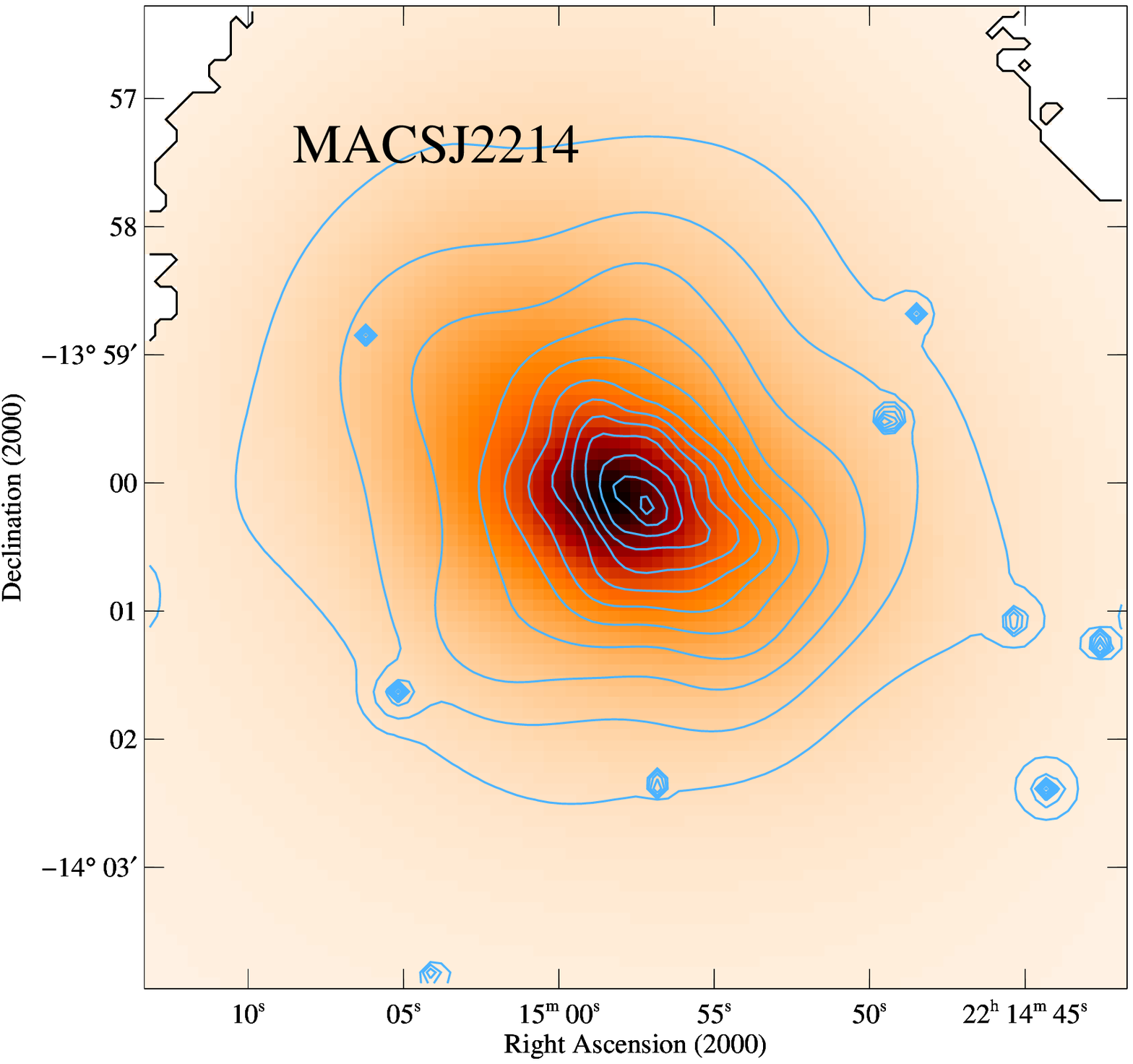} \\
\contcaption{}
\end{figure*}

 \end{document}